\documentclass[sigconf,balance=false]{acmart}

\usepackage{booktabs}                           
\usepackage{subcaption} 
\usepackage[pdf,singlefile]{graphviz}
\usepackage{array}
\usepackage{color}
\usepackage{syntax}
\usepackage{framed}
\usepackage{listings}
\usepackage{pifont}
\usepackage{multirow}

\usepackage{tikz}
\usetikzlibrary{positioning,fit,calc,shapes.geometric}

\usepackage{hhline}

\setcopyright{none}

\begin{document}

\definecolor{darkgreen}{RGB}{0,127,0}
\definecolor{darkred}{RGB}{127,0,0}
\definecolor{shadecolor}{RGB}{250, 250, 250}
\definecolor{lightyellow}{RGB}{250, 250, 180}
\definecolor{HLYELLOW}{RGB}{240, 127, 0}
\definecolor{hlyellow}{RGB}{240, 127, 0}

\lstset{ %
	language=Python,
	frame=single, 
	framerule=0pt,
	basicstyle=\scriptsize\ttfamily, 
	backgroundcolor=\color{shadecolor},
	keywordstyle=\color{blue}\bfseries,
	commentstyle=\color{darkgreen},
	rulecolor=\color{black},
	stringstyle=\color{darkred},
	lineskip=0pt,
	keywords={def, @dace, program, map, lambda, dace, float32, with, as, True, False, return, for, in, class, @staticmethod},
	numbers=left,
	numbersep=5pt,
	showstringspaces=false
}

\newcommand{\one}{\ding{182}}
\newcommand{\two}{\ding{183}}
\newcommand{\three}{\ding{184}}
\newcommand{\four}{\ding{185}}
\newcommand{\five}{\ding{186}}
\newcommand{\six}{\ding{187}}
\newcommand{\seven}{\ding{188}}
\newcommand{\htor}[1]{\textcolor{green}{[htor: #1]}}

\renewcommand{\paragraph}[1]{{\noindent\textbf{\textit{#1}}~~~}}

\setlength{\textfloatsep}{0.5em}

\title[Stateful Dataflow Multigraphs]{Stateful Dataflow Multigraphs: A Data-Centric Model for Performance Portability on Heterogeneous Architectures}

\author[Ben-Nun et al.]{Tal Ben-Nun, Johannes de Fine Licht, Alexandros N. Ziogas, Timo Schneider, Torsten Hoefler}

\affiliation{
  \department{Department of Computer Science, ETH Zurich, Switzerland}    
}
\email{{talbn,definelicht,alziogas,timos,htor}@inf.ethz.ch}

\begin{abstract}
The ubiquity of accelerators in high-performance computing has driven programming complexity beyond the skill-set of the average domain scientist. To maintain performance portability in the future, it is imperative to decouple architecture-specific programming paradigms from the underlying scientific computations.
We present the Stateful DataFlow multiGraph (SDFG), a data-centric intermediate representation that enables separating program definition from its optimization. By combining fine-grained data dependencies with high-level control-flow, SDFGs are both expressive and amenable to program transformations, such as tiling and double-buffering. These transformations are applied to the SDFG in an interactive process, using extensible pattern matching, graph rewriting, and a graphical user interface. We demonstrate SDFGs on CPUs, GPUs, and FPGAs over various motifs --- from fundamental computational kernels to graph analytics. We show that SDFGs deliver competitive performance, allowing domain scientists to develop applications naturally and port them to approach peak hardware performance without modifying the original scientific code.
\end{abstract}


\copyrightyear{2019} 
\acmYear{2019} 
\acmConference[SC '19]{The International Conference for High Performance Computing, Networking, Storage, and Analysis}{November 17--22, 2019}{Denver, CO, USA}
\acmBooktitle{The International Conference for High Performance Computing, Networking, Storage, and Analysis (SC '19), November 17--22, 2019, Denver, CO, USA}
\acmPrice{15.00}
\acmDOI{10.1145/3295500.3356173}
\acmISBN{978-1-4503-6229-0/19/11}

\maketitle

\section{Motivation}

\begin{figure}[t]
	\centering
	\includegraphics[width=.95\linewidth,clip]{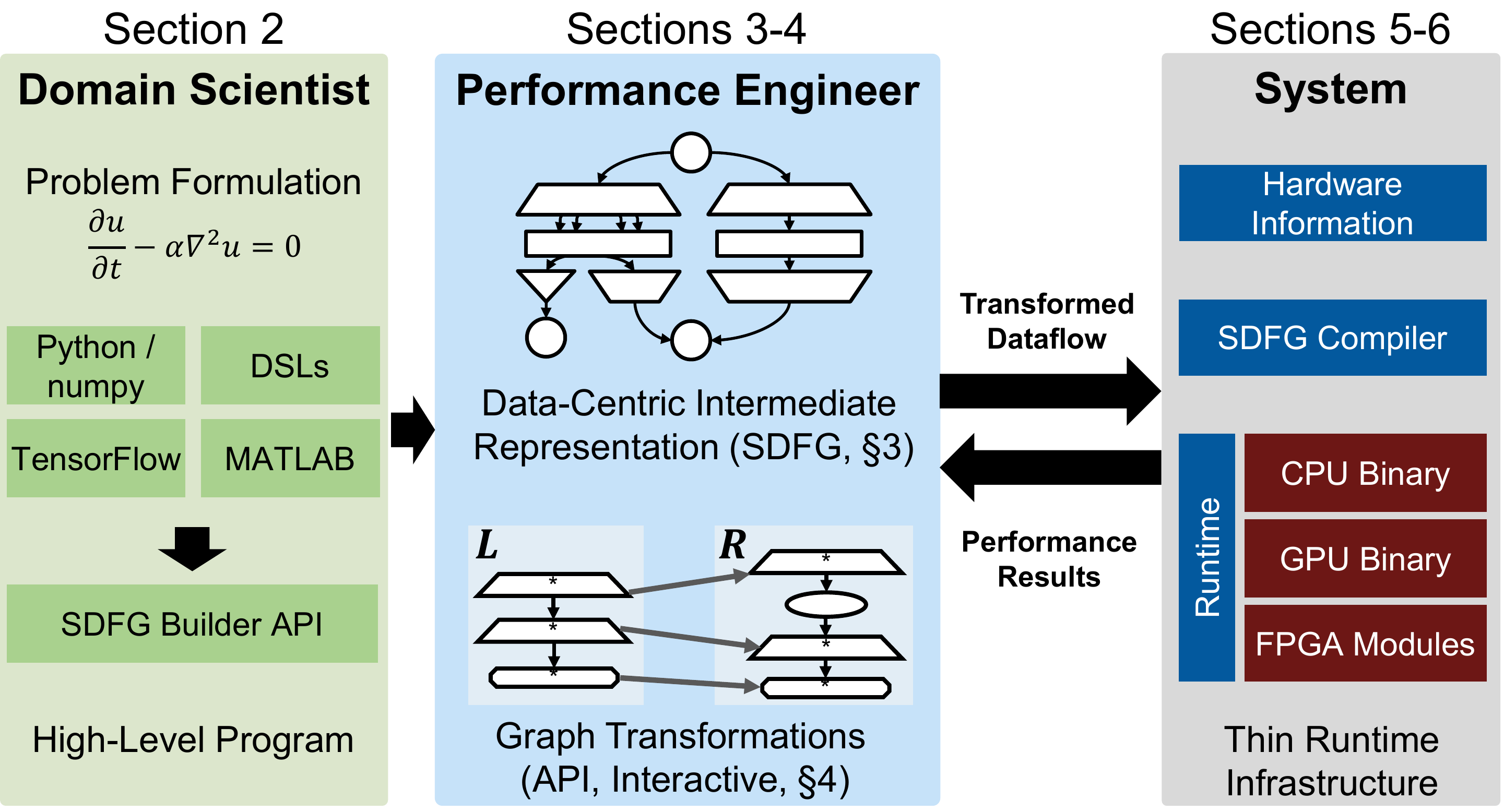}
	\vspace{-1em}
	\caption{Proposed Development Scheme}
	\label{fig:philosophy}
\end{figure}

HPC programmers have long sacrificed ease of programming and portability
for achieving better performance. This mindset was established at a time
when computer nodes had a single processor/core and were programmed with
C/Fortran and MPI. 
The last decade, witnessing the end of Dennard scaling and Moore's law,
brought a flurry of new technologies into the compute nodes. Those range
from simple multi-core and manycore CPUs to heterogeneous GPUs and
specialized FPGAs. 
To support those architectures, the complexity of OpenMP's specification grew by more than
an order of magnitude from 63 pages in OpenMP 1.0 to 666 pages in OpenMP
5.0.
This one example illustrates how (performance) programming complexity
shifted from network scalability to node utilization. 
Programmers would now not only worry about communication (fortunately,
the MPI specification grew by less than 4x from MPI-1.0 to 3.1) but also
about the much more complex on-node heterogeneous programming. 
The sheer number of new approaches, such as OpenACC, OpenCL, or CUDA demonstrate
the difficult situation in on-node programming. 
This increasing complexity makes it nearly impossible for domain
scientists to write portable and performant code today.

The growing complexity in performance programming led to a
specialization of roles into domain scientists and performance
engineers.
Performance engineers typically optimize codes by moving functionality
to performance libraries such as BLAS or LAPACK.
If this is insufficient, they translate 
the user-code to optimized versions, often in different languages such as
assembly code, CUDA, or tuned OpenCL.
Both libraries and manual tuning reduce code maintainability, because the optimized
versions are not only hard to understand for the original author (the
domain scientist) but also cannot be changed without major effort.

Code annotations as used by OpenMP or OpenACC do not change the original
code that then remains understandable to the domain programmer.
However, the annotations must re-state (or modify) some of the semantics
of the annotated code (e.g., data placement or reduction operators).
This means that a (domain scientist) programmer who modifies the code,
\emph{must} modify some annotations or she may introduce hard-to-find
bugs.
With heterogeneous target devices, it now becomes common that the
complexity of annotations is higher than the code they
describe~\cite{pat-maccormicks-sos-talk}.
Thus, scientific programmers can barely manage the complexity of the
code targeted at heterogeneous devices.

The main focus of the community thus moved from scalability to
performance portability as a major research
target~\cite{extreme-heterogeneity}. 
\emph{We call a code-base \textbf{performance-portable} if the domain
scientist's view (``what is computed'') does not change while the code
is optimized to different target architectures, achieving consistently high performance.
The execution should be approximately as performant (e.g., attaining similar ratio of peak performance) as the best-known implementation or theoretical best performance on the target architecture~\cite{perfport}.}
As discussed before, hardly any existing programming model that supports
portability to different accelerators satisfies this definition.

\sloppy
Our Data-centric Parallel Programming (DAPP) concept addresses
performance portability. It uses a data-centric viewpoint of an
application to separate the roles of domain scientist and
performance programmer, as shown in Fig.~\ref{fig:philosophy}.
DAPP relies on Stateful DataFlow multiGraphs (SDFGs) to represent code
semantics and transformations, and supports modifying
them to tune for particular target architectures.
It bases on the observation that data-movement dominates time and
energy in today's computing systems~\cite{padal} and pioneers the
necessary fundamental change of view in parallel programming.
As such, it builds on ideas of data-centric mappers and schedule annotations 
such as Legion~\cite{legion} and Halide~\cite{halide}
\emph{and extends them with a multi-level visualization of data movement, code transformation and compilation for heterogeneous targets,
and strict separation of concerns for programming roles}.
The domain programmer thus works in a convenient and well-known language
such as (restricted) Python or MATLAB. 
The compiler transforms the code into an SDFG, on which the performance engineer
solely works on, specifying transformations that match
certain data-flow structures on all levels (from registers to inter-node communication) and modify them.
Our transformation language can implement arbitrary changes to the SDFG
and supports creating libraries of transformations to optimize
workflows.
Thus, SDFGs separate the concerns of the domain scientist and the
performance engineers through a clearly defined interface, enabling
highest productivity of both roles.

We provide a full implementation of this concept in our Data-Centric 
(DaCe) programming environment, which supports (limited) Python, MATLAB, and
TensorFlow as frontends, as well as support for selected DSLs.
DaCe is easily extensible to other frontends through an SDFG builder
interface.
Performance engineers develop potentially domain-specific \textit{transformation
libraries} (e.g., for stencil-patterns) and can tune them through DaCe's
Interactive Optimization Environment (DIODE).
The current implementation focuses on on-node parallelism as the most
challenging problem in scientific computing today. However, it is
conceivable that the principles can be extended beyond node-boundaries
to support large-scale parallelism using MPI as a backend.

The key contributions of our work are as follows:
\begin{itemize}
  \item We introduce the principle of Data-centric Parallel Programming,
    in which we use Stateful Dataflow Multigraphs, a data-centric
    Intermediate Representation that enables separating code definition
    from its optimization.
  \item We provide an open-source implementation\footnote{\url{https://www.github.com/spcl/dace}} of the data-centric environment and its 
    performance-optimization IDE.
  \item We demonstrate performance portability on fundamental kernels, graph algorithms, and a real-world quantum transport simulator --- results are
    competitive with and faster than expert-tuned libraries from Intel and NVIDIA,
    approaching peak hardware performance, and up to five orders of
    magnitude faster than na\"{i}ve FPGA code written with High-Level
    Synthesis, all from \emph{the same program source code}.
\end{itemize}

\begin{figure}[t]
	\begin{subfigure}[b]{.51\linewidth} 	
	\begin{lstlisting}[numbers=none]
@dace.program
def Laplace(A: dace.float64[2,N], 
            T: dace.uint32):
  for t in range(T):
    for i in dace.map[1:N-1]:
      A[(t+1)%2, i] = \
        A[t%2, i-1:i+2] * [1,-2,1]

a = numpy.random.rand(2, 2033)
Laplace(A=a, T=500)\end{lstlisting}
	\caption{Python Representation}
	\vspace{-1em}
\label{lst:mm:numpy}
	\end{subfigure}\hfill
	\begin{subfigure}[b]{.44\linewidth}
		\includegraphics[width=\linewidth,page=1]{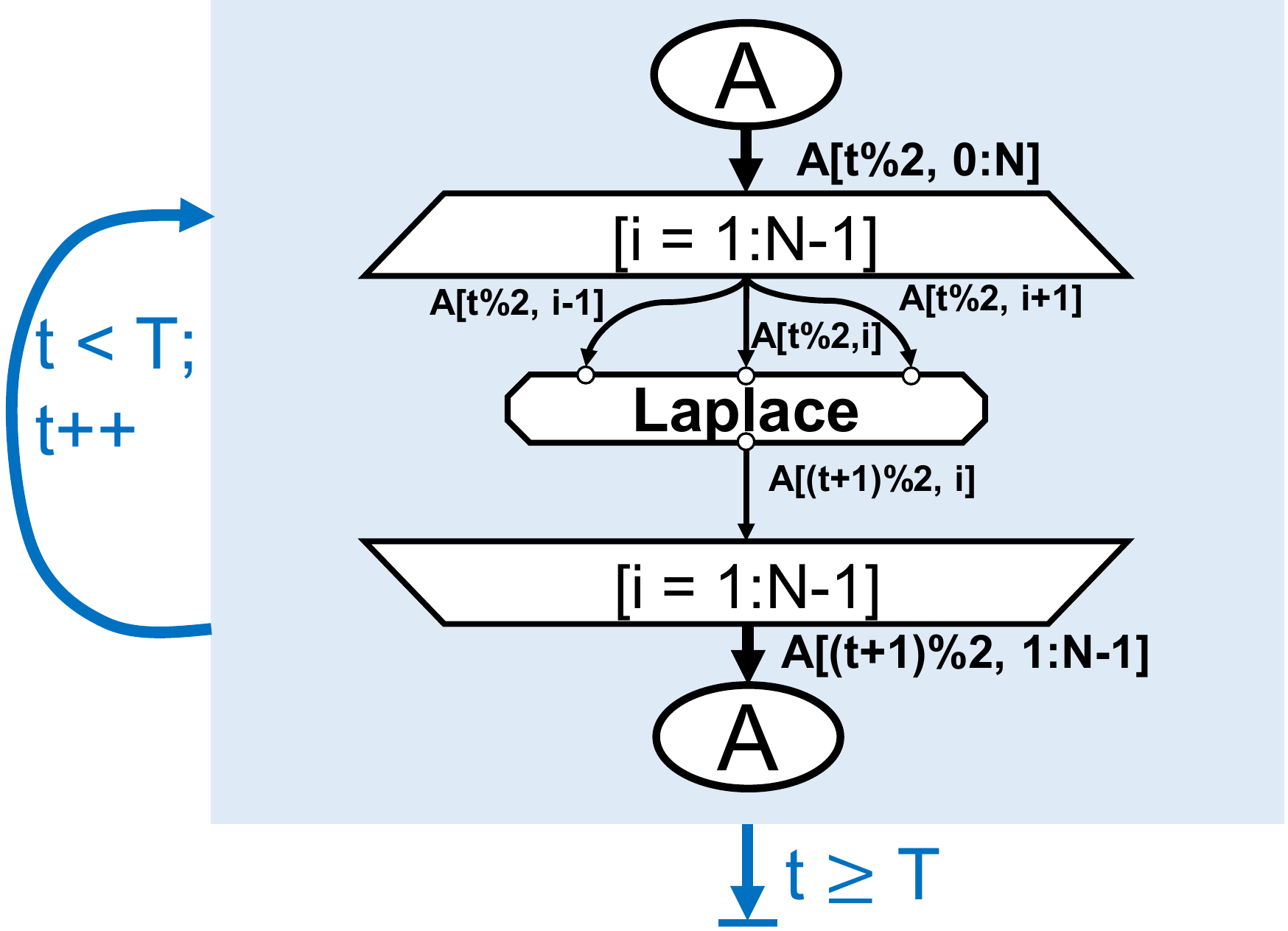}
		\caption{Resulting SDFG}
		\vspace{-1em}		
		\label{fig:mm:sdfg}
	\end{subfigure}
	\caption{Data-Centric Computation of a Laplace Operator}
	\label{lst:mm}
\end{figure}

\section{Data-Centric Programming}
\label{sec:bg}

Current approaches in high-performance computing optimizations \cite{padal}
revolve around improving data locality. Regardless of the underlying
architecture, the objective is to keep information as close as possible
to the processing elements and promote memory reuse. Even a simple
application, such as matrix multiplication, requires multiple stages of
transformations, including data layout modifications (packing) and
register-aware caching \cite{goto08anatomy}. 
Because optimizations do not modify computations and differ for each
architecture, maintaining performance portability of scientific
applications requires separating computational semantics from data
movement.

SDFGs enable separating application development into two stages, as
shown in Fig. \ref{lst:mm}. The problem is formulated as a
high-level program (Fig.~\ref{lst:mm:numpy}), and is then transformed into a human-readable
SDFG as an Intermediate Representation (IR, Fig.~\ref{fig:mm:sdfg}). The SDFG can then
be modified without changing the original code, and as long as the dataflow aspects do not change, the original code can be updated while keeping SDFG transformations intact. What differentiates the SDFG from other IRs is the ability to hierarchically and parametrically view data movement, where scopes in the graph contain overall data requirements. This enables reusing transformations (e.g., tiling) at different levels of the memory hierarchy, as well as performing cross-level optimizations.

\textit{The modifications to the SDFG are \textbf{not} completely
automatic.} Rather, they are made by the performance engineer as a
result of informed decisions based on the program structure, hardware
information, and intermediate performance results. 
To support this, a transformation interface and 
common optimization libraries
should be at the performance engineer's disposal,
enabling modification of the IR in a verifiable manner (i.e., without
breaking semantics), either programmatically or interactively.  
The domain scientist, in turn, writes an entire
application once for all architectures, and can freely update the
underlying calculations without undoing optimizations on the SDFG.

Conceptually, we perform the separation of computation from data movement logic by viewing programs as data flowing between operations, much like Dataflow and Flow-Based Programming \cite{dataflow}.
One key difference between dataflow and data-centric parallel programming,
however, is that in a pure dataflow model execution is stateless, which
means that constructs such as loops have to be unrolled.
At the other extreme, traditional, control-centric programs revolve
around statements that are executed in order. 
Data-centric parallel programming promotes the use of \textit{stateful
	dataflow}, in which execution order depends first on data dependencies,
but also on a global execution state. The former fosters the expression
of concurrency, whereas the latter increases expressiveness and
compactness by enabling concepts such as loops and data-dependent
execution.
The resulting concurrency works in several granularities, from
utilizing processing elements on the same chip, to ensuring overlapped
copy and execution of programs on accelerators in clusters.
A data-centric model combines the following concepts:
\begin{enumerate}
	\item \textbf{Separating Containers from Computation}: Data-holding constructs with volatile or non-volatile information are defined as separate entities from computations, which consist of stateless functional units that perform arithmetic or logical operations in any granularity. 
	\item \textbf{Dataflow}: The concept of information moving from one container or computation to another. This may be translated to copying, communication, or other forms of movement.
	\item \textbf{States}: Constructs that provide a mechanism to introduce execution order independent of data movement.
	\item \textbf{Coarsening}: The ability to view parallel patterns in a hierarchical manner, e.g., by grouping repeating computations.
\end{enumerate}

The resulting programming interface should thus enable these concepts without drastically modifying development process, both in terms of languages and integration with existing codebases.

\subsection{Domain Scientist Interface}
\label{sec:frontends}

\paragraph{Languages}
Scientific applications typically employ different programming models
and Domain-Specific Languages (DSLs) to solve problems. To cater to the
versatile needs of the domain scientists, SDFGs should be easily
generated from various languages. We thus implement SDFG frontends in
high-level languages (Python, MATLAB, TensorFlow), and provide a
low-level (builder) API to easily map other DSLs to SDFGs. In the rest of this section, we focus on the Python \cite{python} interface, which is the most extensible.

\paragraph{Interface}
The Python interface creates SDFGs from restricted Python code, supporting numpy operators and functions, as well as the option to explicitly specify dataflow. In Fig.~\ref{lst:mm:numpy}, we demonstrate the data-centric interface on a one-dimensional Laplace operator.
DaCe programs exist as decorated, strongly-typed functions in the application ecosystem, so that they can interact with existing codes using array-based interfaces (bottom of figure).
The Python interface contains primitives such as \texttt{map} and \texttt{reduce} (which translate directly into SDFG components), allows programmers to use multi-dimensional arrays, and implements an extensible subset of operators from numpy \cite{numpy} on such arrays to ease the use of linear algebra operators. For instance, the code \texttt{A @ B} generates the dataflow of a matrix multiplication.

\paragraph{Extensibility}
For operators and functions that are not implemented, a user can easily provide dataflow implementations using decorated functions (\texttt{@dace.replaces('numpy.conj')}) that describe the SDFG. Otherwise, unimplemented functions fall-back into Python, casting the array pointers (which may be defined internally in the DaCe program) into numpy arrays and emitting a ``potential slowdown'' warning. If the syntax is unsupported (e.g., dynamic dictionaries), an error is raised.

\paragraph{Explicit Dataflow}
If the programmer does not use predefined operators (e.g., for custom element-wise computation), dataflow ``intrinsics'' can be explicitly defined separately from code, in constructs which we call \textit{Tasklets}. Specifically, tasklet functions cannot access data unless it was explicitly moved in or out using pre-declared operators (\texttt{<<}, \texttt{>>}) on arrays, as shown in the code.

\begin{figure}[t]
	\includegraphics[width=0.7\linewidth,trim={0cm 0cm 0cm -0.1cm},page=10]{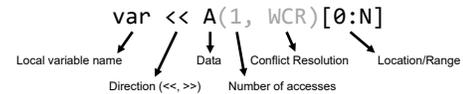}
	\vspace{-1em}
	\caption{Anatomy of a Python Memlet}
	\label{fig:memlet}
	\vspace{-0.5em}
\end{figure}

Data movement operations (\textit{memlets}) can be versatile, and the Python syntax of explicit memlets is defined using the syntax shown in Fig.~\ref{fig:memlet}. First, a local variable (i.e., that can be used in computation) is defined, whether it is an input or an output. After the direction of the movement, the data container is specified, along with an optional range (or index). In some applications (e.g., with indirect or data-dependent access), it is a common occurrence that the \textit{subset} of the accessed data is known, but not exact indices; specifying memory access constraints both enables this behavior and facilitates access tracking for decomposition, e.g., which data to send to an accelerator. Finally, the two optional values in parentheses govern the nature of the access --- the number of data elements moved, used for performance modeling, and a lambda function that is called when write-conflicts may occur. 

\begin{figure}[h]
	\vspace{-1em}
	\begin{lstlisting}[numbers=none]
@dace.program
def spmv(A_row: dace.uint32[H + 1], A_col: dace.uint32[nnz], 
         A_val: dace.float32[nnz], x: dace.float32[W], 
         b: dace.float32[H]):
  for i in dace.map[0:H]:
    for j in dace.map[A_row[i]:A_row[i+1]]:
      with dace.tasklet:
        a << A_val[j]
        in_x << x[A_col[j]]
        out >> b(1, dace.sum)[i]    
        out = a * in_x\end{lstlisting}
    \vspace{-1.5em}
	\caption{Sparse Matrix-Vector Mult. with Memlets}
	\vspace{-1em}
	\label{fig:indirect}
\end{figure}

Using explicit dataflow is beneficial when defining nontrivial data accesses. Fig.~\ref{fig:indirect} depicts a full implementation of Sparse Matrix-Vector multiplication (SpMV). In the implementation, the access \texttt{x[A\_col[j]]} is translated into an indirect access subgraph (see Appendix~F) that can be identified and used in transformations.

\begin{figure}[b]
	\begin{lstlisting}[numbers=none]
@dace.program
def extmm(A: dace.complex128[M,K], B: dace.complex128[K,N],
          C: dace.complex128[M,N]):
  with dace.tasklet(language=dace.Language.CPP,
                    code_global='#include <mkl.h>'):
    a << A; b << B; in_c << C; out_c >> C
    '''
    dace::complex128 alpha(1, 0), beta(0, 0);
    cblas_zgemm(CblasRowMajor, 'N', 'N', M, N, K, &alpha, a, M,
                b, K, &beta, out_c, M);
    '''\end{lstlisting}
	\vspace{-1.5em}
	\caption{External Code in DaCe}
	\label{fig:extcode}
\end{figure}

\paragraph{External Code}
Supporting scientific code, in terms of performance and productivity, requires the ability to call previously-defined functions or invoke custom code (e.g., intrinsics or assembly). In addition to falling back to Python, the frontend enables defining tasklet code in the generated code language directly. In Fig.~\ref{fig:extcode} we see a DaCe program that calls a BLAS function directly. The semantics of such tasklets require that memlets are defined separately (for correctness); the code can in turn interact with the memory directly (memlets that are larger than one element are pointers). With this feature, users can use existing codes and benefit from concurrent scheduling that the SDFG provides.

\paragraph{Parametric Dimensions}
To support parametric sizes (e.g., of arrays and maps) in DaCe, we utilize symbolic math evaluation. In particular, we extend the SymPy \cite{sympy} library to support our expressions and strong typing. The code can thus define symbolic sizes and use complex memlet subset expressions, which will be analyzed during SDFG compilation.
The \textit{separation of access and computation, flexible interface, and symbolic sizes} are the core enablers of data-centric parallel programming, helping domain scientists create programs that are amenable to efficient hardware mapping.

\begin{table}[t]
	\caption{SDFG Syntax}
	\vspace{-1em}
	\label{tbl:prims}
	\small
	\begin{tabular}{|>{\centering\arraybackslash}m{0.5\linewidth}|b{0.4\linewidth}|}
		
		\hline
		\bf Primitive & \bf Description \\\hline
\multicolumn{2}{|l|}{\bf Data-Centric Model}\\\hline
		\vspace{-1em}\digraph[height=0.3in,trim={1.25cm 1.3cm 1.25cm 1cm},clip]{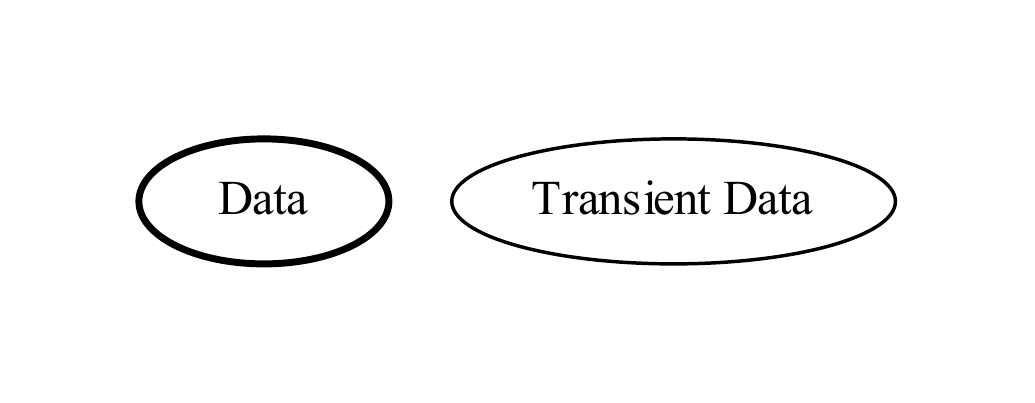}{subgraph { rank=same; "Data" [width=1, style="bold"]; "Transient Data" [width=1]}}& \textbf{Data}: N-dimensional array container. \\\hline
		\vspace{-1em}\digraph[height=0.3in,trim={1.25cm 1.4cm 1.25cm 1cm},clip]{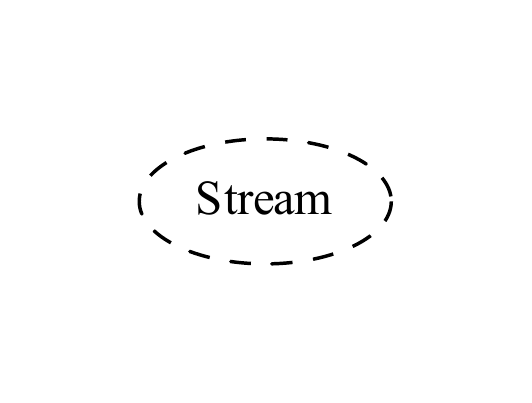}{"Stream" [style=dashed, width=1]} & \textbf{Stream}: Streaming data container. \\\hline		
		\vspace{-1em}\digraph[height=0.3in,trim={1.25cm 1.4cm 1.25cm 1cm},clip]{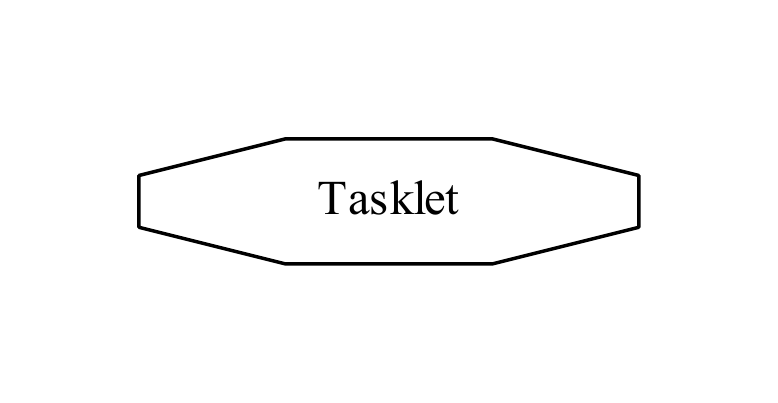}{"Tasklet" [shape=octagon, width=2]} & \textbf{Tasklet}: Fine-grained computational block.\\\hline
		\digraph[height=0.4in,trim={1.8cm 1.6cm 3.2cm 1cm},clip]{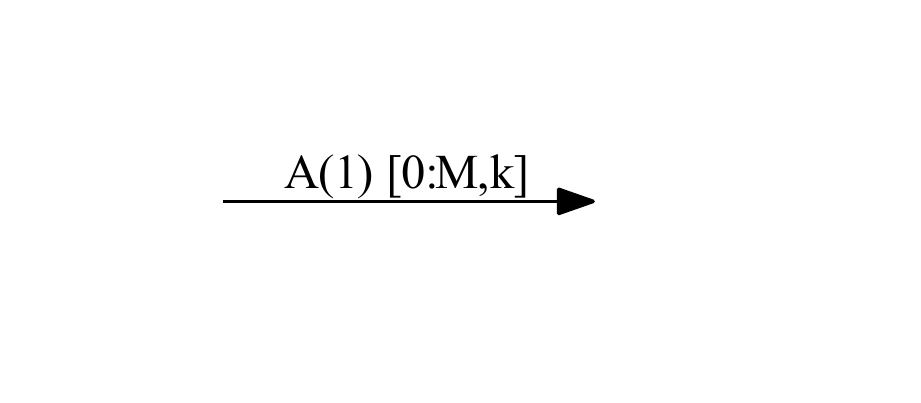}{rankdir=LR;  b [style=invis, width=0, height=0, shape=rectangle]; c[style=invis]; b->c [label="A(1) [0:M,k]"]} & \textbf{Memlet}: Data movement descriptor.\\\hline
		\digraph[height=0.3in,trim={1.4cm 1.4cm 1.4cm 1cm},clip]{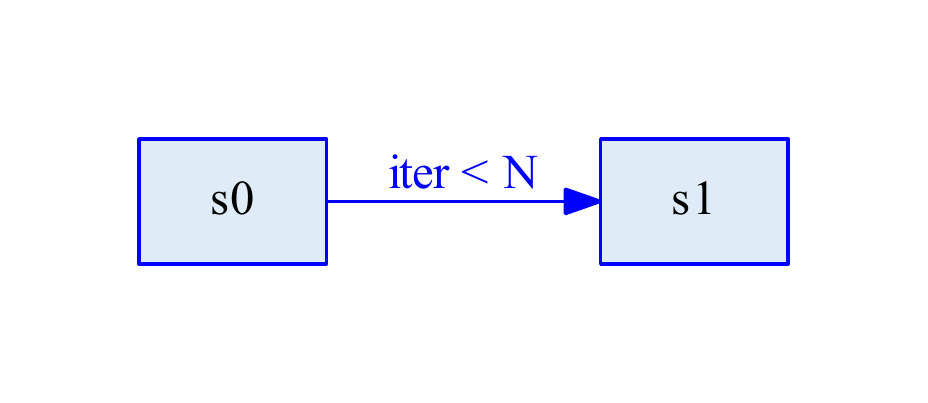}{rankdir=LR; s0 [shape=rectangle, color=blue, fillcolor="0.58, 0.1, 0.97", style=filled];s1 [shape=rectangle, color=blue, fillcolor="0.58, 0.1, 0.97", style=filled]; s0->s1 [label="iter < N", color=blue, fontcolor=blue]} & 
		\textbf{State}: State machine element. \\\hline
\multicolumn{2}{|l|}{\bf Parametric Concurrency}\\\hline
		\vspace{-0.5em}\digraph[width=\linewidth,trim={1.25cm 1.4cm 1.25cm 1cm},clip]{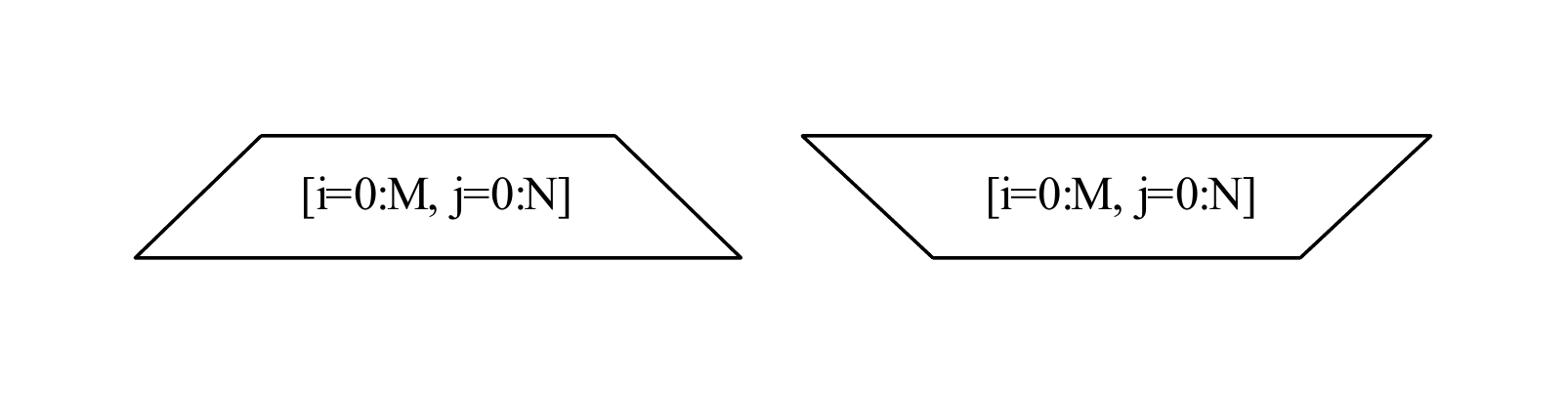}{subgraph { rank=same; "[i=0:M, j=0:N]" [shape=trapezium]; " [i=0:M, j=0:N]" [shape=invtrapezium]}}\vspace{-1em} & \textbf{Map}: Parametric graph abstraction for parallelism.\\\hline
		\vspace{-0.8em}\digraph[width=\linewidth,trim={1.25cm 1.4cm 1.25cm 1cm},clip]{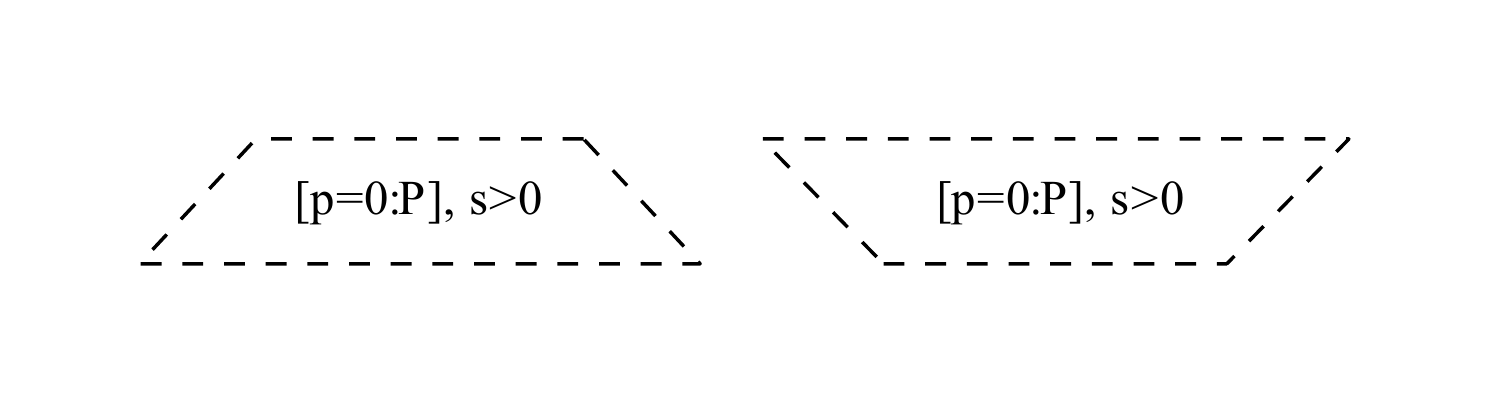}{subgraph { rank=same; "[p=0:P], s>0" [shape=trapezium style=dashed width=2]; " [p=0:P], s>0" [shape=invtrapezium style=dashed width=2]}}\vspace{-1em} & \textbf{Consume}: Dynamic mapping of computations on streams. \\\hline
		\vspace{-0.3in}\digraph[height=0.3in,trim={1.8cm 1.6cm 3.2cm 1.25cm},clip]{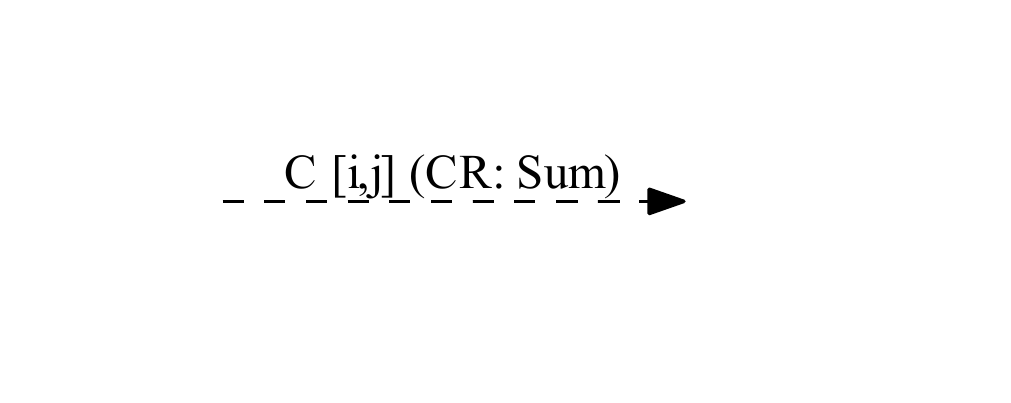}{rankdir=LR;  b [style=invis, width=0, height=0, shape=rectangle]; c[style=invis]; b->c [label="C [i,j] (CR: Sum)", style=dashed]} & \textbf{Write-Conflict Resolution}: Defines behavior during conflicting writes.\\\hline
\multicolumn{2}{|l|}{\bf Parallel Primitives and Nesting}\\\hline
		\vspace{-0.8em}\digraph[height=0.22in,trim={1.4cm 1.4cm 1.4cm 1.8cm},clip]{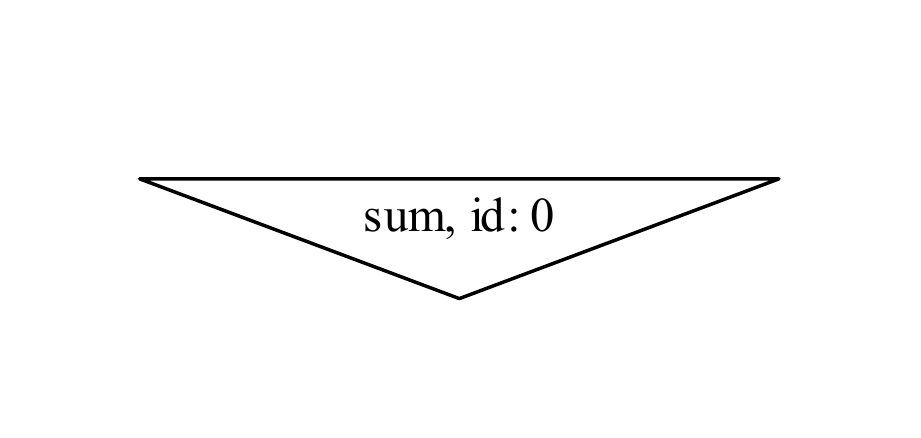}{"sum, id: 0" [shape=invtriangle]} & \textbf{Reduce}: Reduction over one or more axes.\\\hline
		\digraph[height=0.3in,trim={1.25cm 1.4cm 1.25cm 1cm},clip]{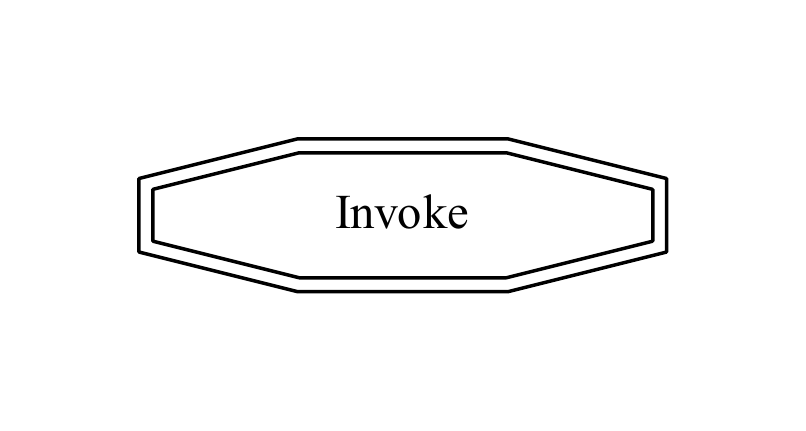}{"Invoke" [shape=doubleoctagon, width=2]} & \textbf{Invoke}: Call a nested SDFG.\\\hline
	\end{tabular}
\end{table}

\section{Stateful Dataflow Multigraphs}
\label{sec:sdfg}

We define an SDFG as a \textit{directed graph of directed acyclic
multigraphs}, whose components are summarized in Table \ref{tbl:prims}.
Briefly, the SDFG is composed of acyclic dataflow multigraphs,
in which nodes represent \textit{containers} or \textit{computation}, and edges (memlets) represent \textit{data movement}.
To support cyclic data dependencies and control-flow, these multigraphs reside in \textit{\textbf{State}} nodes at the top-level graph. Following complete execution of the dataflow in a state, state transition edges on the top-level graph specify conditions and assignments, forming a state machine. For complete operational semantics of SDFGs, we refer to Appendix A.

\subsection{Containers}
As a data-centric model, SDFGs offer two forms of data containers: \textbf{Data} and \textbf{Stream} nodes. 
\textbf{Data} nodes represent a location in memory that is mapped to a multi-dimensional array, whereas \textbf{Stream} nodes are defined as multi-dimensional arrays of concurrent queues, which can be accessed using push/pop semantics.
Containers are tied to a specific \textit{storage location} (as a node property), which may be on a GPU or even a file. In the generated code, memlets between containers either generate appropriate memory copy operations or fail with illegal accesses (for instance, when trying to access paged CPU memory within a GPU kernel).
In FPGAs, Stream nodes instantiate FIFO interfaces that can be used to connect hardware modules.
Another property of containers is whether they are \textit{transient}, i.e., only allocated for the duration of SDFG execution. This allows transformations and performance engineers to distinguish between buffers that interact with external systems, and ones that can be manipulated (e.g., data layout) or eliminated entirely, even across devices. This feature is advantageous, as standard compilers cannot make this distinction, especially in the presence of accelerators.

\begin{figure}[t]
	\begin{subfigure}[b]{.25\linewidth}
		\includegraphics[width=\linewidth,page=2]{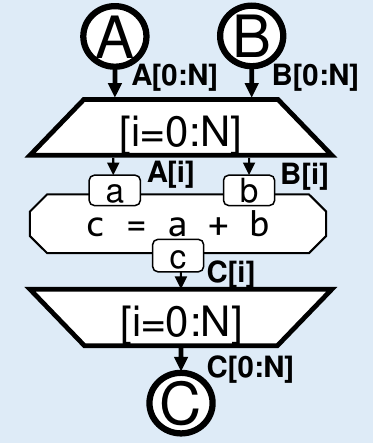}
		\caption{Parametric}
		\label{fig:map:param}
	\end{subfigure}
	\hfil
	\begin{subfigure}[b]{.57\linewidth}
		\includegraphics[width=\linewidth,page=3]{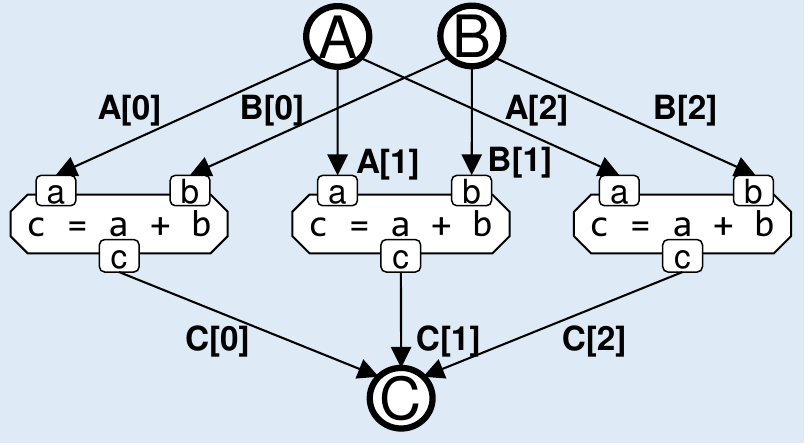}
		\caption{Expanded ($N=3$)}
		\label{fig:map:expanded}
	\end{subfigure}
	\vspace{-1em}
	\caption{Parametric Parallelism in SDFGs}
	\label{fig:map}
\end{figure}

\subsection{Computation}
\textbf{Tasklet} nodes contain stateless, arbitrary computational functions of any granularity. 
The SDFG is designed, however, for fine-grained tasklets, so as to enable performance engineers to analyze and optimize the most out of the code, leaving computational semantics intact.
Throughout the process of data-centric transformations and compilation, the tasklet code remains immutable.
This code, provided that it cannot access external memory without memlets, can be written in any source language that can compile to the target platform, and is implemented in Python by default.

In order to support Python as a high-level language for tasklets, we implement a Python-to-C++ converter. The converter traverses the Python Abstract Syntax Tree (AST), performs type and shape inference, tracks local variables for definitions, and uses features from C++14 (such as lambda expressions and \texttt{std::tuple}s) to create the corresponding code.
Features that are not supported include dictionaries, dynamically-sized lists, exceptions, and other Python high-level constructs.
Given that tasklets are meant to be fine-grained, and that our DaCe interface is strongly typed (\S~\ref{sec:frontends}), this feature-set is sufficient for HPC kernels and real-world applications.

\subsection{Concurrency}

Expressing parallelism is inherent in SDFGs by design, supported by the \textbf{Map} and \textbf{Consume} scopes.
Extending the traditional task-based model, SDFGs expose concurrency by grouping parallel subgraphs (computations, local data, movement) into one symbolic instance, enclosed within two ``scope'' nodes.
Formally, we define an enclosed subgraph as nodes dominated by a scope entry node and post-dominated by an exit node.
The subgraphs are thus connected to external data only through scope nodes, which enables analysis of their overall data requirements (useful, e.g., for automatically transforming a map to GPU code).

\begin{figure}[t]
	\begin{subfigure}[b]{.45\linewidth}
		\centering
		\includegraphics[height=1.05in,page=1]{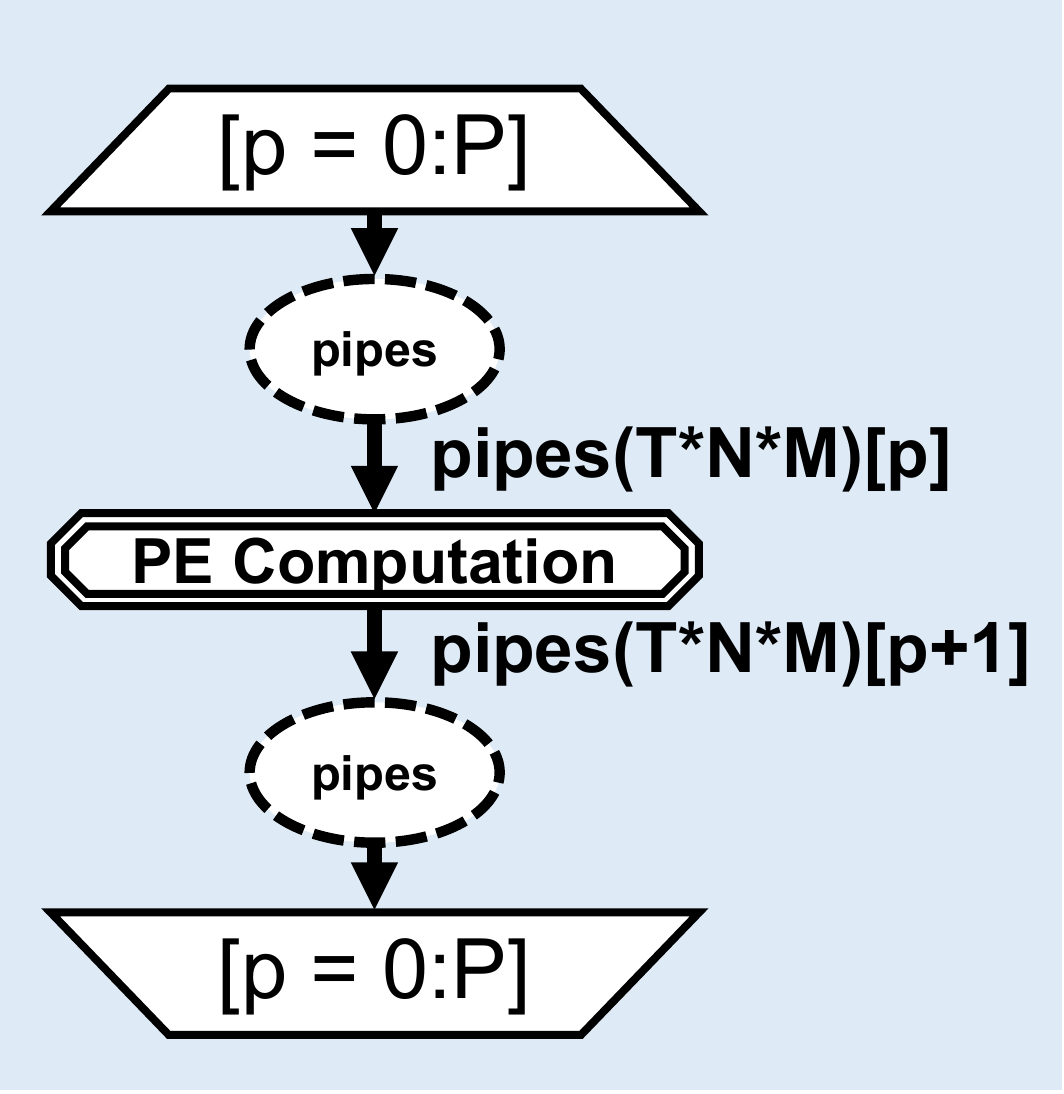}
		\caption{SDFG}
	\end{subfigure}
	\begin{subfigure}[b]{.45\linewidth}
		\centering
		\includegraphics[height=1.05in,page=2]{figures/systolic.pdf}
		\caption{Hardware Modules}
	\end{subfigure}
	\vspace{-1em}
	\caption{Parametric Generation of Systolic Arrays}
	\label{fig:systolicarray}
\end{figure}

\textbf{Map} scopes represent parallel computation on all levels, and can be nested hierarchically. This feature consolidates many parallel programming concepts, including multi-threading, GPU kernels, multi-GPU synchronization, and multiple processing elements on FPGAs.
The semantics of a Map are illustrated in Fig.~\ref{fig:map} --- a symbolic integer set attribute of the scope entry/exit nodes called \textit{range} (Fig.~\ref{fig:map:param}) defines how the subgraph should be expanded on evaluation (Fig.~\ref{fig:map:expanded}).
Like containers, Maps are tied to \textit{schedule}s that determine how they translate to code. When mapped to multi-core CPUs, Map scopes generate OpenMP \texttt{parallel for} loops; for GPUs, device schedules generate CUDA kernels (with the map range as thread-block indices), whereas thread-block schedules determine the dimensions of blocks, emitting synchronization calls (\texttt{\_\_syncthreads}) as necessary; for FPGAs, Maps synthesize different hardware modules as processing elements.
Streams can also be used in conjunction with Maps to compactly represent systolic arrays, constructs commonly used in circuit design to represent efficient pipelines, as can be seen in Fig.~\ref{fig:systolicarray}. Note that no data is flowing in or out of the Map scope (using empty memlets for the enclosed subgraph) --- this would replicate the scope's contents as separate connected components.


\begin{figure}[h]
	\centering
	\vspace{-0.75em}
	\includegraphics[width=.8\linewidth,page=3]{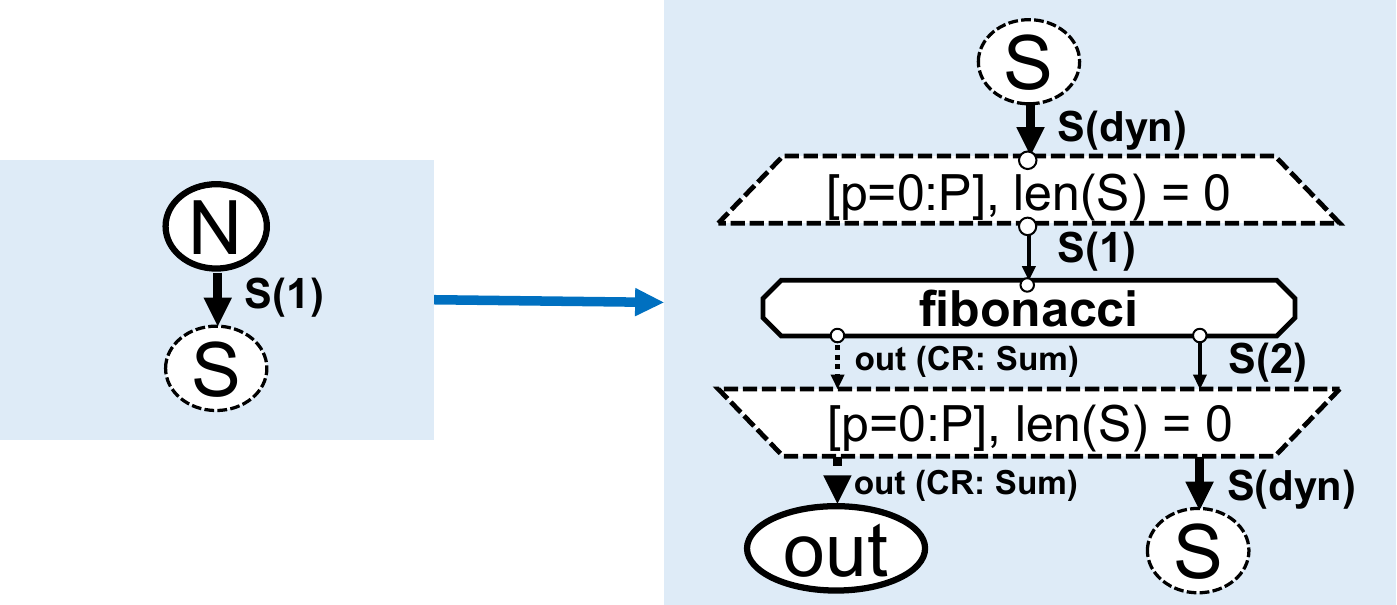}
	\vspace{-1em}
  \caption{Asynchronous Fibonacci SDFG}
	\label{fig:sdfg-fib}
	\vspace{-1em}	
\end{figure}

\textbf{Consume} scopes enable producer/consumer relationships via dynamic processing of streams. Consume nodes are defined by the \textit{number of processing elements}, an \textit{input stream} to consume from, and a \textit{quiescence condition} that, when evaluated to true, stops processing. An example is shown in Fig.~\ref{fig:sdfg-fib}, which computes the Fibonacci recurrence relation of an input $N$ without memoization. In the SDFG, the value is first pushed into the stream $S$ and asynchronously processed by $P$ workers, with the memlet annotated as \texttt{dyn} for dynamic number of accesses. The tasklet adds the result to \texttt{out} and pushes two more values to $S$ for processing. The consume scope then operates until the number of elements in the stream is zero, which terminates the program.

Consume scopes are implemented using batch stream dequeue and atomic operations to asynchronously pop and process elements. The potential to encompass complex parallel patterns like work stealing schedulers using high-performance implementations of this node dramatically reduces code complexity.

\begin{figure}[t]
	\centering
	\vspace{1em}
	\begin{subfigure}[b]{.45\linewidth}	
		\centering
		\includegraphics[height=1.3in,page=4]{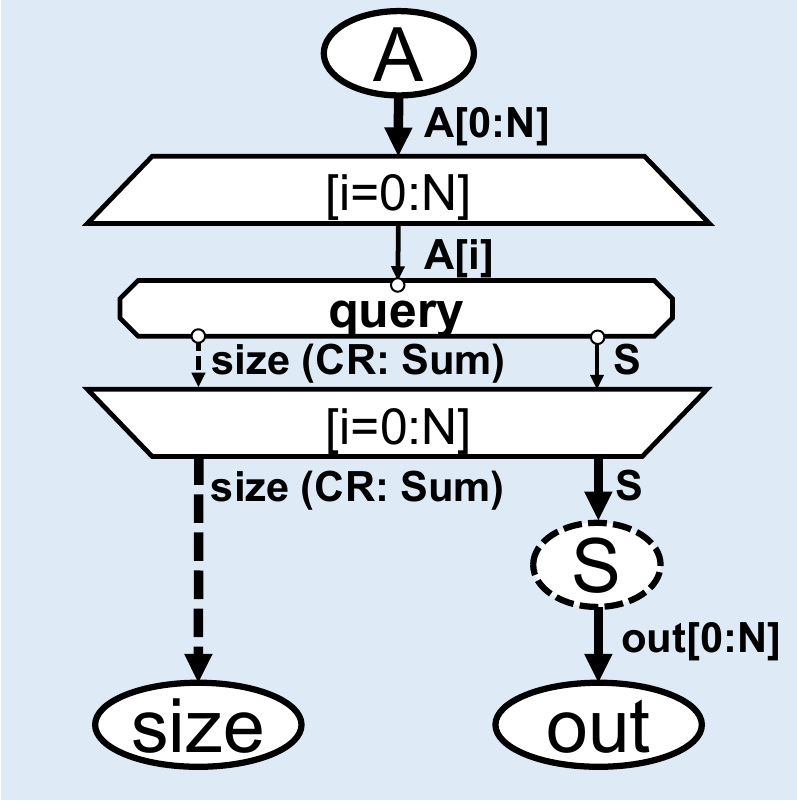}
		\caption{Query}		
		\label{fig:sdfg-pbf}
	\end{subfigure}		
	\quad
	\begin{subfigure}[b]{.45\linewidth}
		\centering
		\includegraphics[height=1.3in,page=2]{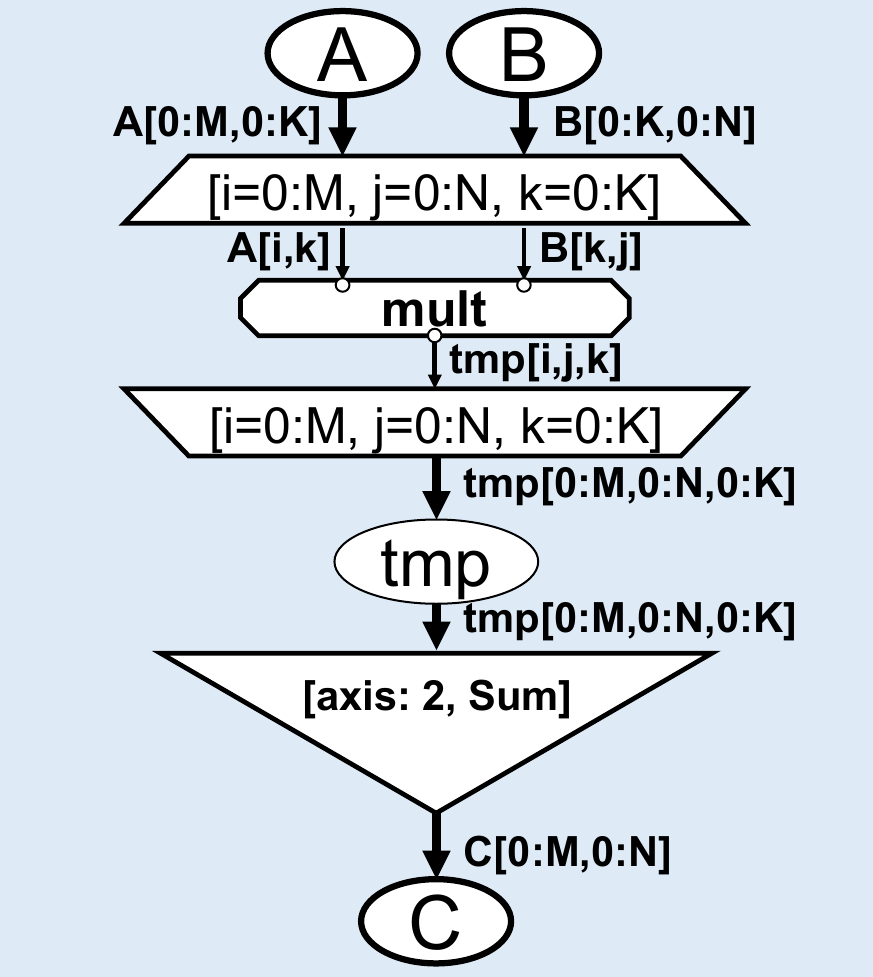}
		\caption{Matrix Multiplication}
		\label{fig:sdfg-mm}
	\end{subfigure}

	\vspace{-1em}
	\caption{Write-Conflicts and Reductions}
\end{figure}

In order to handle concurrent memory writes from scopes, we define \textbf{Write-Conflict Resolution} memlets. As shown in Fig.~\ref{fig:sdfg-pbf}, such memlets are visually highlighted for the performance engineer using dashed lines. Implementation-wise, such memlets can be implemented as atomic operations, critical sections, or accumulator modules, depending on the target architecture and the function.
\textbf{Reduce} nodes complement conflict resolution by implementing target-optimized reduction procedures on data nodes. An example can be seen with a map-reduce implementation of matrix multiplication (Fig.~\ref{fig:sdfg-mm}), where a tensor with multiplied pairs of the input matrices is reduced to the resulting matrix. As we shall show in the next section, this inefficient representation can be easily optimized using data-centric transformations.

Different connected components within an SDFG multigraph also run concurrently (by definition).
Thus, they are mapped to \texttt{parallel sections} in OpenMP, different CUDA streams on GPUs, or different command queues on FPGAs. 
These concepts are notoriously cumbersome to program manually for all platforms, where synchronization mistakes, order of library calls, or less-known features (e.g., \texttt{nowait}, non-blocking CUDA streams) may drastically impact performance or produce wrong results. 
Therefore, the SDFG's automatic management of concurrency, and configurable fine-tuning of synchronization aspects by the performance engineer (e.g., maximum number of concurrent streams, nested parallelism) make the IR attractive for HPC programming on all platforms.

\subsection{State}

\begin{figure}[t]
	\centering
	\vspace{1em}
	\begin{subfigure}[b]{.5\linewidth}	
		\includegraphics[height=1.15in,page=1]{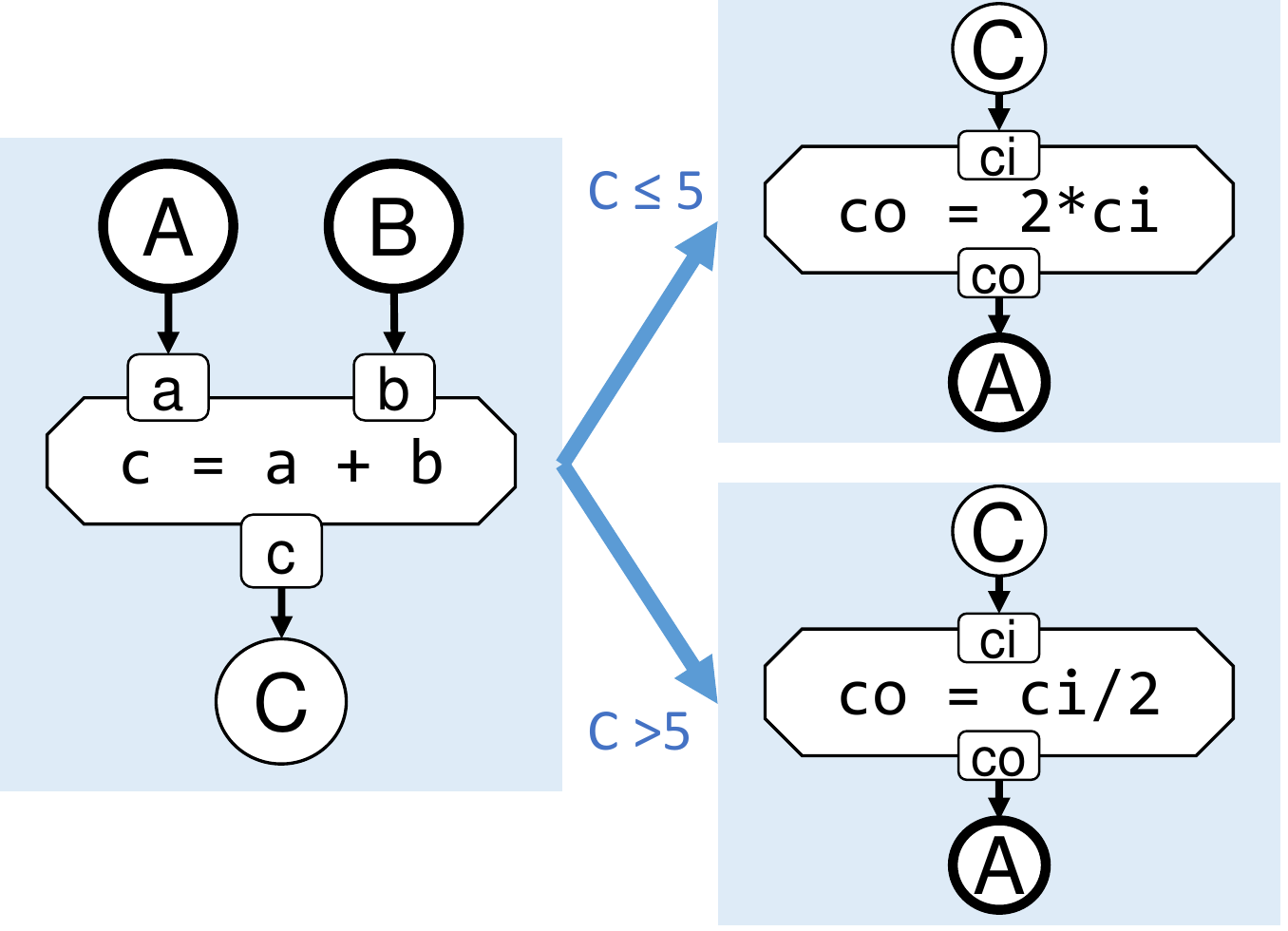}
		\caption{Branching}
		\label{fig:simplesdfg}
	\end{subfigure}		
	\quad
	\begin{subfigure}[b]{.45\linewidth}
		\centering
		\includegraphics[width=.9\linewidth]{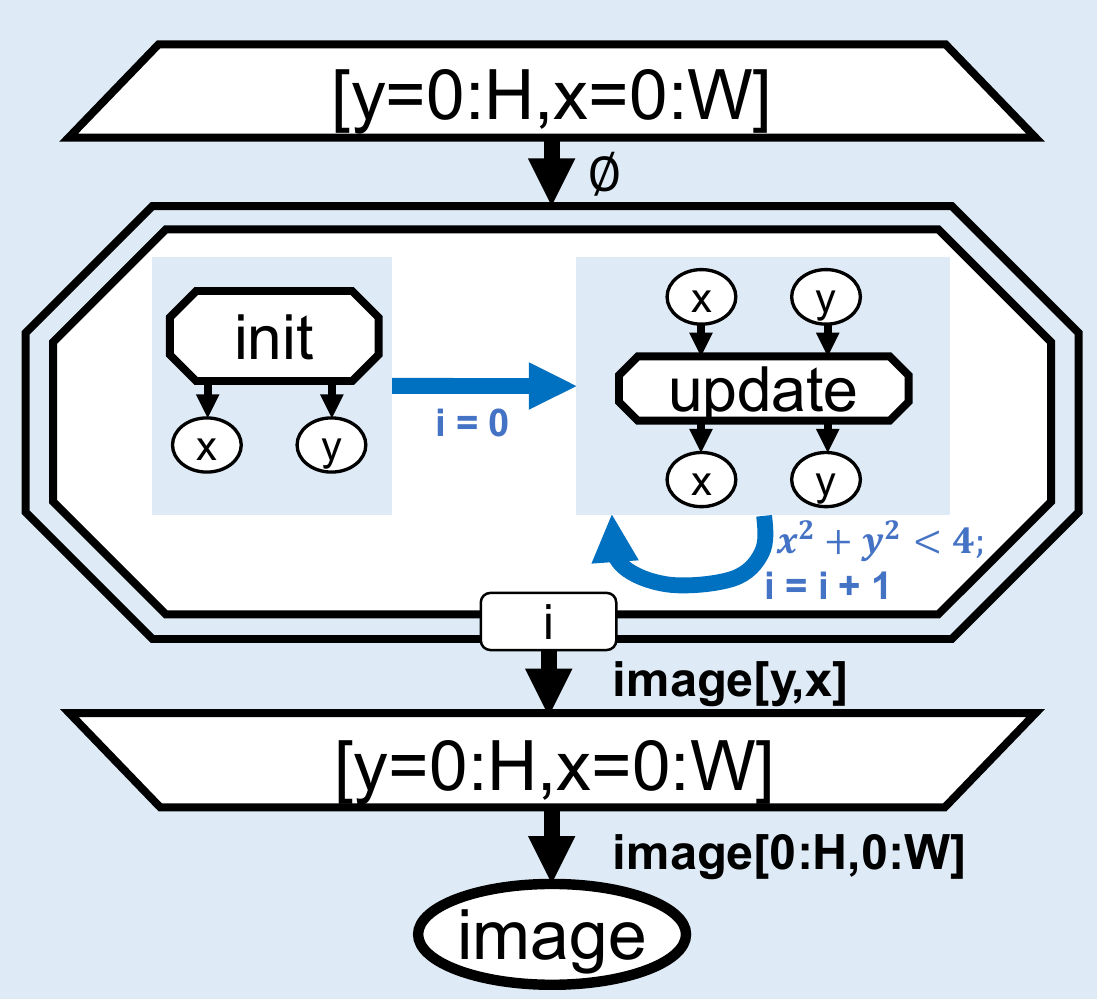}
		\caption{Mandelbrot (Nested SDFG)}
		\label{fig:mandelbrot}
	\end{subfigure}
	\vspace{-2em}
	\caption{Data-Dependent Execution}
\end{figure}

Sequential operation in SDFGs either implicitly occurs following data dependencies, or explicitly specified using multiple states. 
State transition edges define a condition, which can depend on data in containers, and a list of assignments to inter-state symbols (e.g., loop iteration). 
The concept of a state machine enables both complex control flow patterns, such as imperfectly nested loops, and data-dependent execution, as shown in Fig.~\ref{fig:simplesdfg}.

To enable control-flow within data-flow (e.g., a loop in a map), or a parametric number of state machines, SDFGs can be nested via the \textbf{Invoke} node. 
The semantics of Invoke are equivalent to a tasklet, thereby disallowing access to external memory without memlets.
The Mandelbrot example (Fig.~\ref{fig:mandelbrot}) demonstrates nested SDFGs. In the program, each pixel requires a different number of iterations to converge. In this case, an invoke node calls another SDFG within the map to manage the convergence loop. 
Recursive calls to the same SDFG are disallowed, as the nested state machine may be inlined or transformed by the performance engineer.

\section{Performance Engineer Workflow}

The Stateful Dataflow Multigraph is designed to expose application data movement motifs, regardless of the underlying computations.
As such, the optimization process of an SDFG consists of finding and leveraging such motifs, in order to mutate program dataflow.
Below, we describe the two principal tools we provide the performance engineer to guide optimization, followed by the process of compiling an SDFG into an executable library.

\subsection{Graph Transformations}
\label{sec:transform}

Informally, we define a graph transformation on an SDFG as a ``find and
replace'' operation, either within one state or between several, which
can be performed if all of the conditions match. For general
optimization strategies (e.g., tiling), we provide a standard library of
such transformations, which is meant to be used as a baseline for
performance engineers. Transformations can be designed with symbolic expressions, or specialized for given sizes in order to fine-tune applications. A list of 16 transformations implemented in DaCe can be found in Appendix~B. 

\sloppy
Transformations consist of a \textit{pattern} subgraph and a \textit{replacement} subgraph. 
A transformation also includes a \textit{matching} function,
used to identify instances of the pattern in an SDFG, and programmatically verify that requirements are met. To find matching subgraphs in SDFGs, we use the VF2 algorithm \cite{vf2} to find isomorphic subgraphs.
Transformations can be applied interactively, or using a Python API for matching and applying transformations. An example of a full source code of a transformation is found in Appendix~D.
Using the transformation infrastructure, we enable seamless knowledge transfer of optimizations across applications.

\begin{figure}[t]
	\centering
	\begin{subfigure}[t]{\linewidth}
		\centering
		\includegraphics[width=.7\linewidth,page=6]{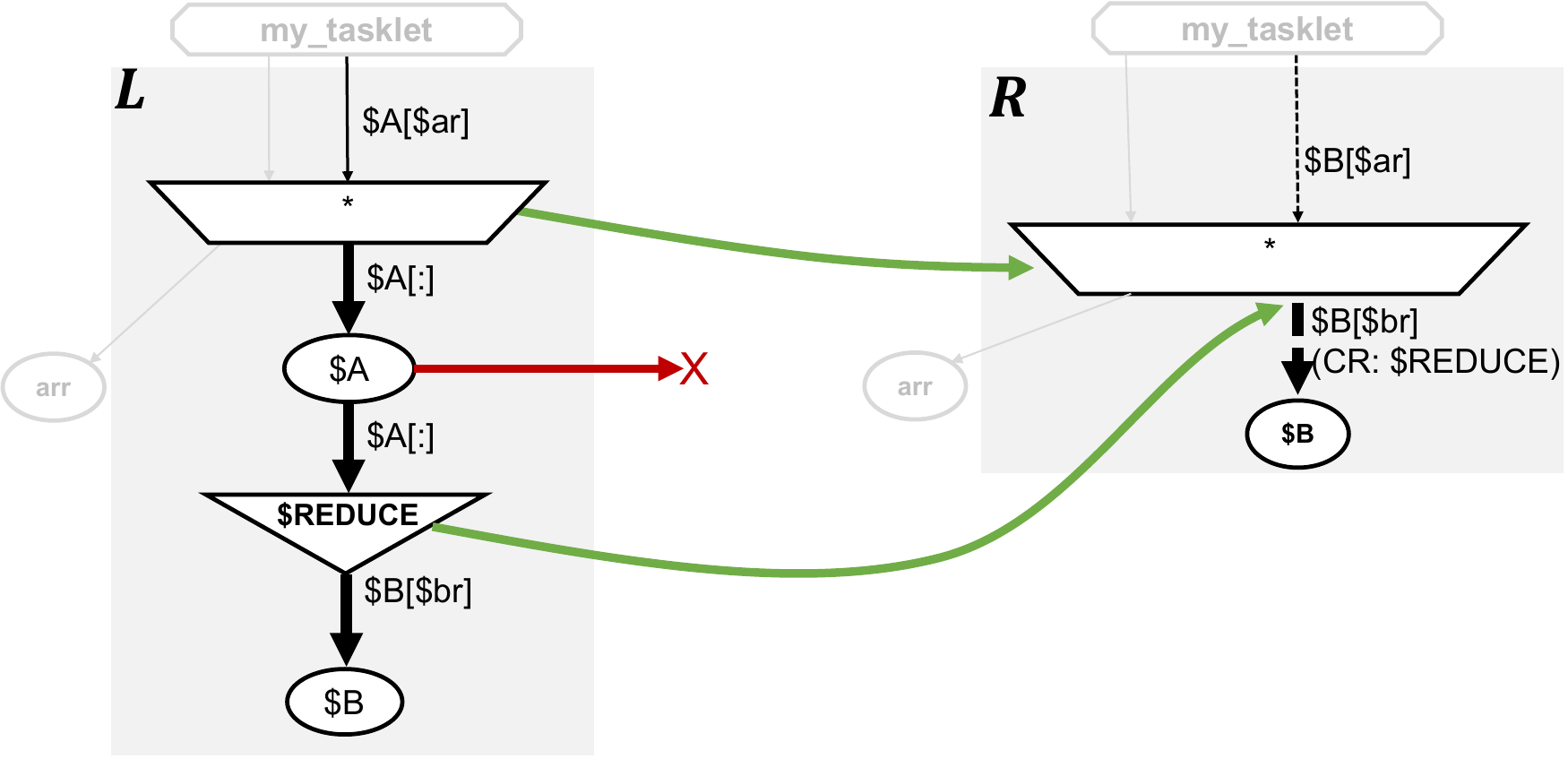}
		\caption{Map-Reduce Fusion}
		\label{fig:unmapreduce}
	\end{subfigure}\\
	\begin{subfigure}[t]{\linewidth}
		\centering
		\includegraphics[width=.7\linewidth,page=7]{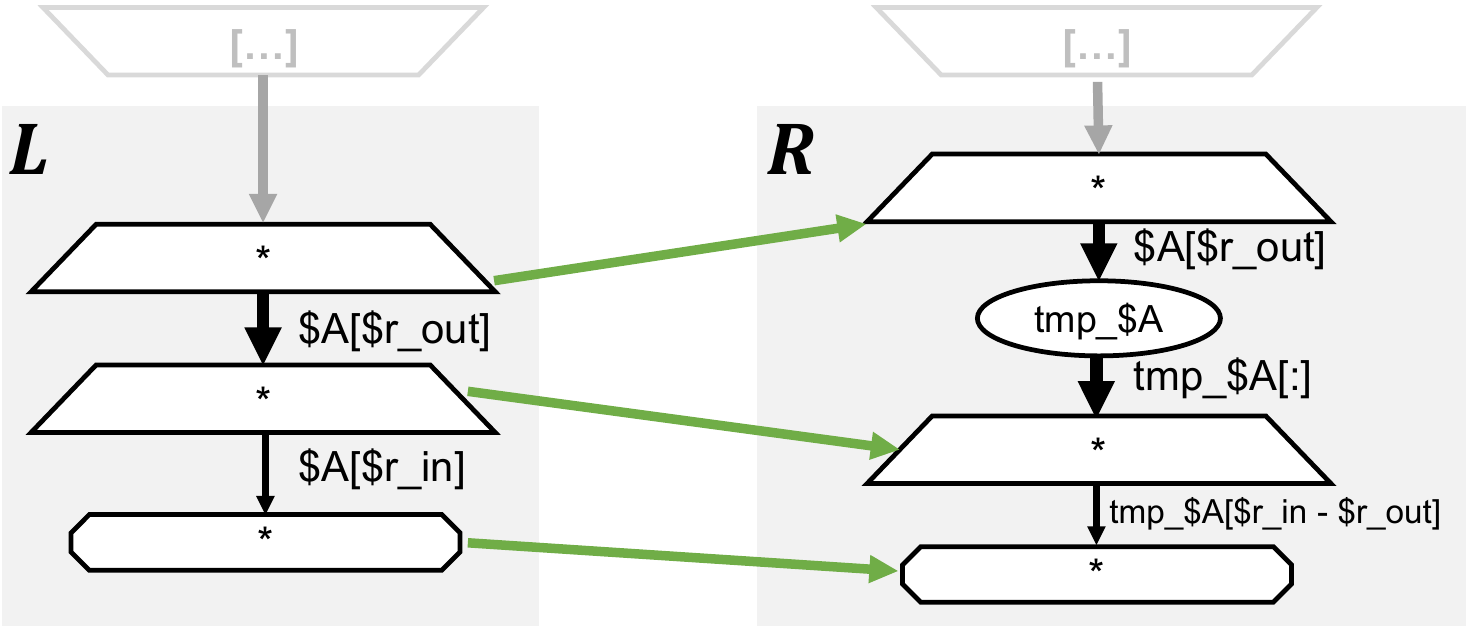}
		\caption{Local Storage}
		\label{fig:localstorage}
	\end{subfigure}
	\vspace{-1em}
	\caption{SDFG Transformations}
	\vspace{-1em}
\end{figure}

Two examples of SDFG transformations can be found in Figures \ref{fig:unmapreduce} and \ref{fig:localstorage}. In Fig. \ref{fig:unmapreduce}, the transformation detects a pattern $L$ where Reduce is invoked immediately following a Map, reusing the computed values. In this case, the transient array \texttt{\$A} can be removed (if unused later) and computations can be fused with a conflict resolution, resulting in the replacement $R$. 
In the second example (Fig. \ref{fig:localstorage}), a local array, which can be assigned to scratch-pad memory, is added between two map nodes. As a result, the relative indices are changed in all subsequent memlets.

\subsection{DIODE}

SDFGs are intended to be malleable and interactive, which we realize with an Integrated Development Environment (IDE). The Data-centric Integrated Optimization Development Environment, or \textbf{DIODE} (Fig.~\ref{fig:diode}), is a specialized IDE for performance engineers that enables editing SDFGs and applying transformations in a guided manner, in addition to the programmatic interface. In particular, performance engineers can:

\begin{figure}[t]
	\centering
	\includegraphics[width=.9\linewidth]{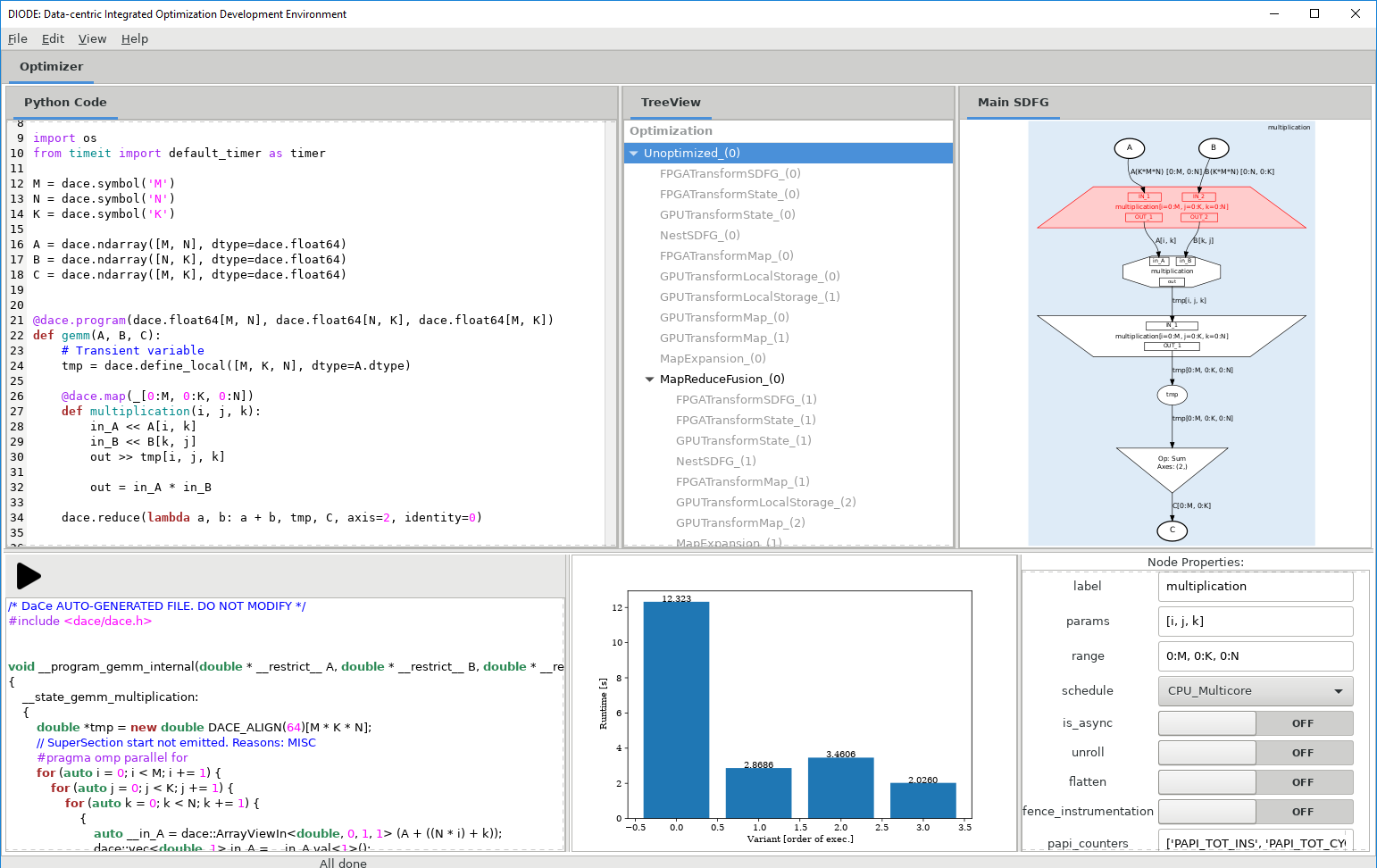}
	\vspace{-1em}
	\caption{DIODE Graphical User Interface}
	\label{fig:diode}
\end{figure}
\begin{figure*}[t]
	\begin{subfigure}{\linewidth}
		\includegraphics[width=\linewidth]{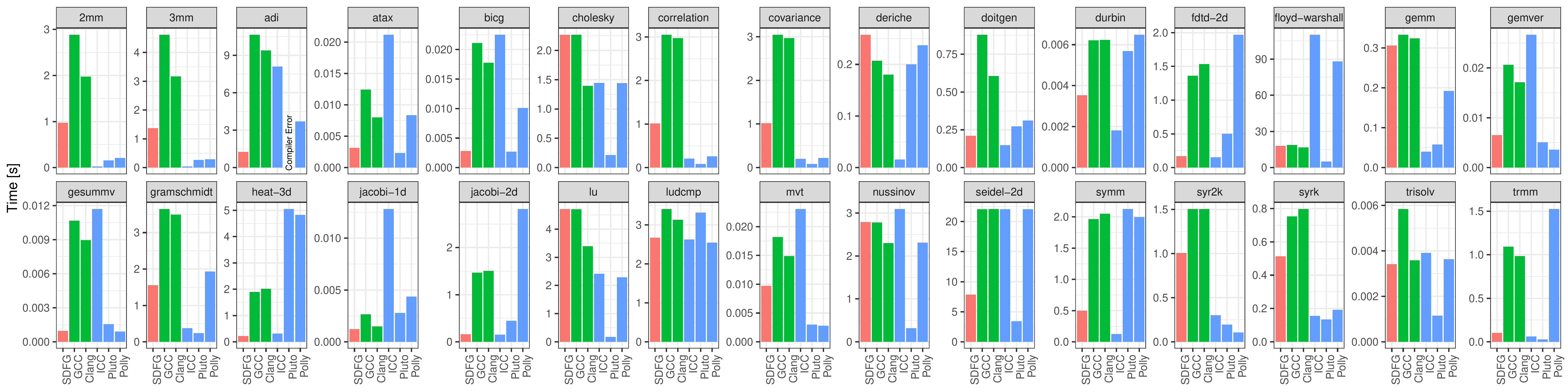}
		\caption{CPU (1.43$\times$ geometric mean speedup over best general-purpose compiler)}
	\end{subfigure}
	\begin{subfigure}{.48\linewidth}
		\includegraphics[width=\linewidth]{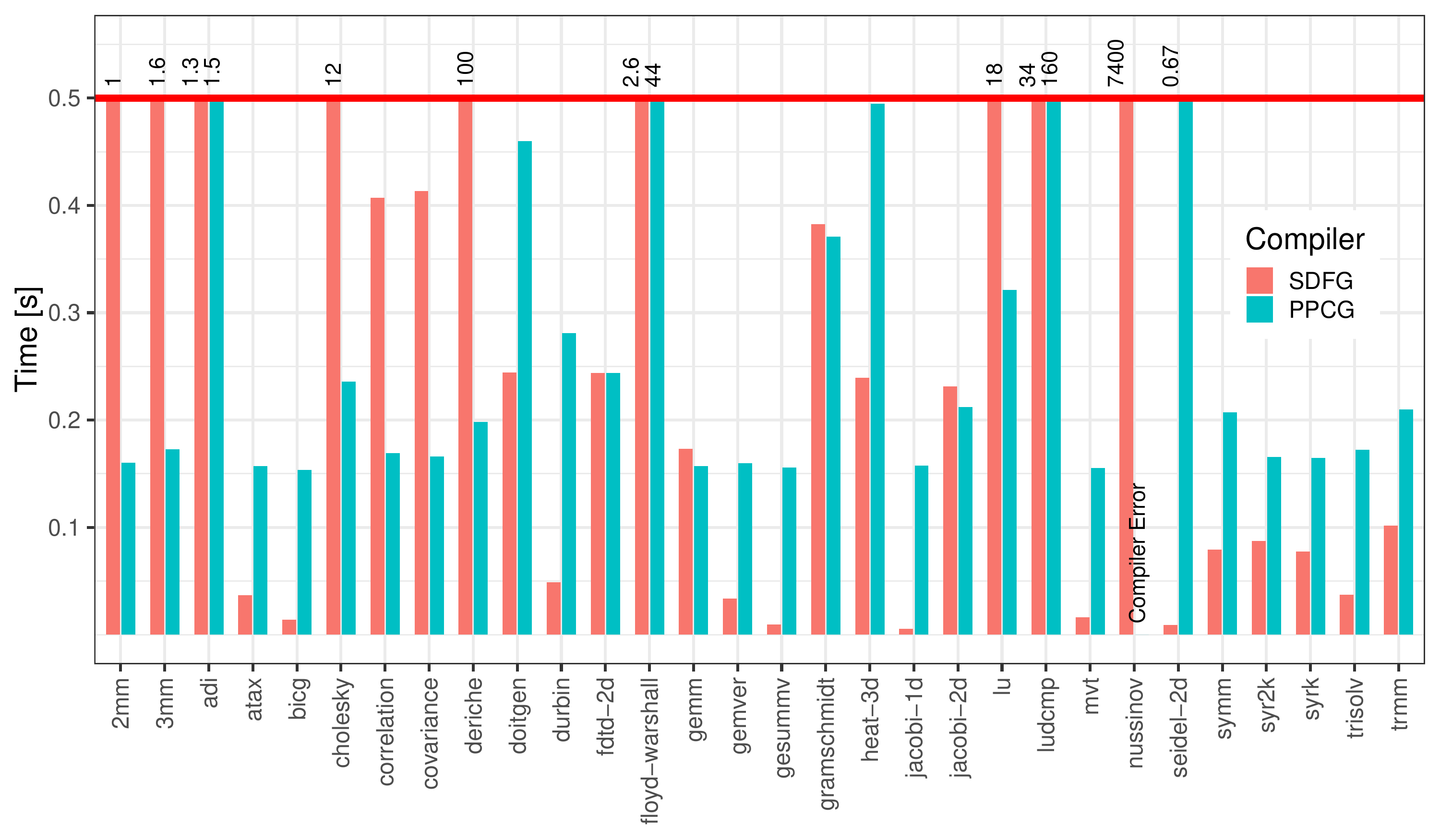}
		\caption{GPU (1.12$\times$ geometric mean speedup over PPCG)}
	\end{subfigure}
	\hfill
	\begin{subfigure}{.48\linewidth}
		\includegraphics[width=\linewidth]{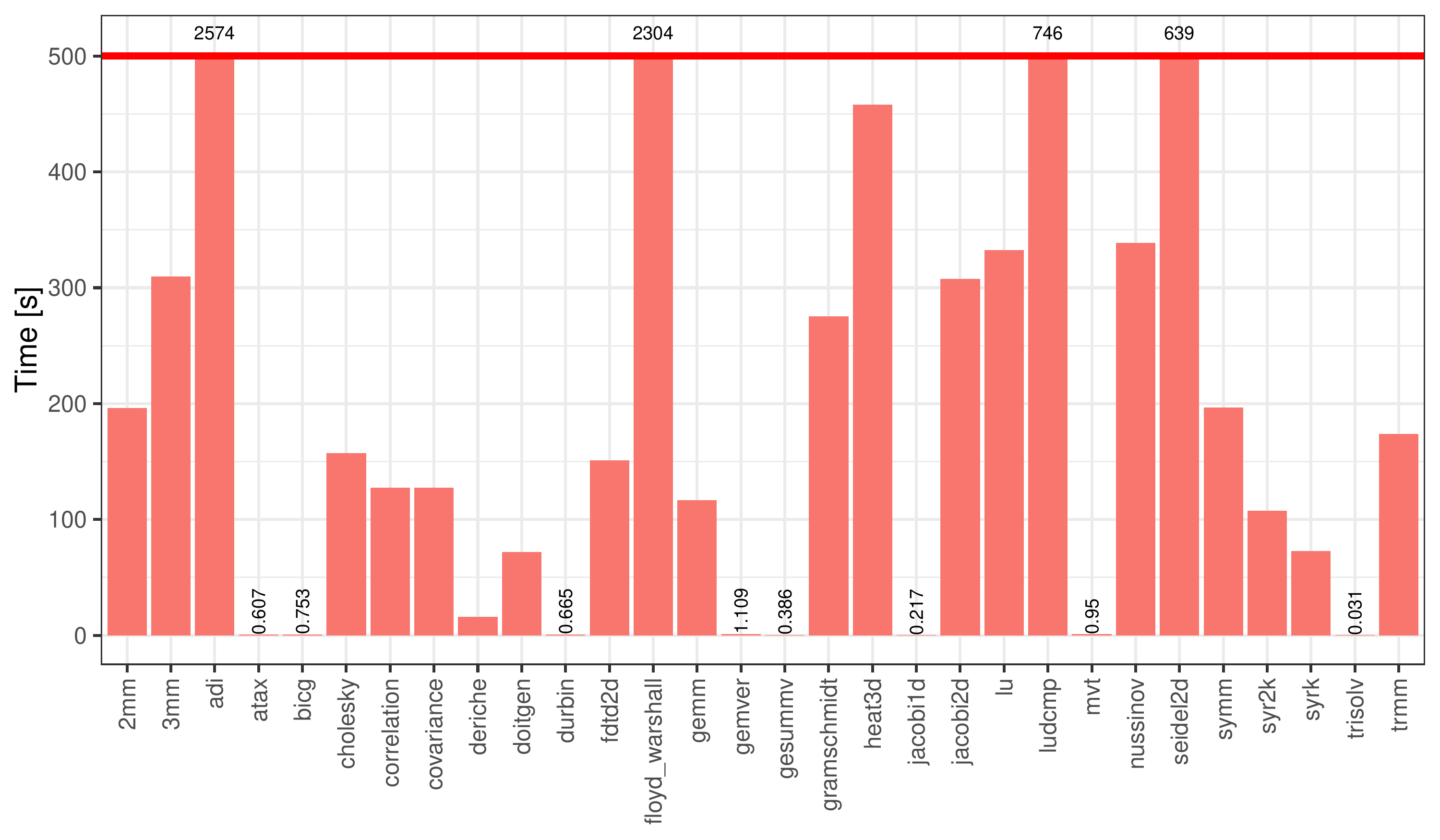}
		\caption{FPGA}
	\end{subfigure}
	\vspace{-1em}
	\caption{Polyhedral Application Runtime (lower is better, best viewed in color)}
	\vspace{-1em}
	\label{fig:polybench}
\end{figure*}

\begin{itemize}
	\item interactively modify attributes of SDFG nodes and memlets;
	\item transform and tune transformation parameters;
	\item inspect an integrated program view that maps between lines in the domain code, SDFG components, and generated code;
	\item run and compare historical performance of transformations;
	\item save transformation chains to files, a form of ``optimization version control'' that can be used when tuning to different architectures (diverging from a mid-point in the chain); and
	\item hierarchically interact with large-scale programs, collapsing irrelevant parts and  focusing on bottleneck subgraphs.
\end{itemize}

We demonstrate the process of interactively optimizing the matrix multiplication SDFG (Fig. \ref{fig:sdfg-mm}) in a video\footnote{\url{https://www.vimeo.com/301317247}}. Using the IDE and the SDFG representation yields nearly the same performance as Intel's optimized Math Kernel Library (MKL) \cite{mkl} in minutes (\S\ref{sec:perf:mm}).

\subsection{Compilation Pipeline}

Compiling an SDFG consists of three steps: \one~ data dependency inference, \two~ code generation, and \three~ compiler invocation.

In step \one, data dependencies on all levels are resolved. First, a validation pass is run on the graph to ensure that scopes are correctly structured, memlets are connected properly, and map schedules and data storage locations are feasible (failing when, e.g., FPGA code is specified in a GPU map). 
Then, memlet ranges are propagated from tasklets and containers outwards (through scopes) to obtain the overall data dependencies of each scope, using the image of the scope function (e.g., Map range) on the union of the internal memlet subsets.
This information is later used to generate exact memory copy calls to/from accelerators.

The code generation process of an SDFG (step \two) is hierarchical, starting from top-level states and traversing into scopes.
It begins by emitting external interface code and the top-level state machine. Within each state, nodes are traversed in topological order, and a platform-specific dispatcher is assigned to generate the respective code, depending on the node's storage or schedule type. The process continues recursively via map/consume scopes and nested SDFGs, generating heterogeneous codes using several dispatchers. Between states, transitions are generated by emitting for-loops and branches when detected, or using conditional \texttt{goto} statements as a fallback. 

In step \three, we invoke compiler(s) for the generated code according to the used dispatchers. 
The compiled library can then be used directly by way of inclusion in existing applications, or through Python/DaCe.

\section{Assessing Performance without Transformations}
\label{sec:perf:poly}

To understand how the inherently-concurrent representation of the SDFG creates reasonably performing na\"{i}ve code, 
we reimplement and run the Polybench \cite{polybench} benchmark over SDFGs, \textit{without any optimizing transformations}, using the experimental setup of Section~\ref{sec:perf}. We show that the representation itself exposes enough parallelism to compete with state-of-the-art polyhedral compilers, outperform them on GPUs, and provide \textit{\textbf{the first complete set of placed-and-routed Polybench kernels on an FPGA}}.

To demonstrate the wide coverage provided by SDFGs, we
apply the \texttt{FPGATransform} automatic transformation to offload each Polybench application to the FPGA during runtime, use our simulation flow to verify correctness, and finally
perform the full placement and routing process. The same applies for \texttt{GPUTransform}. We execute all kernels on the accelerators, including potentially unprofitable ones (e.g., including tasklets without parallelism).

The results are shown in Fig. \ref{fig:polybench}, comparing SDFGs both with general-purpose compilers (green bars in the figure), and with pattern-matching and polyhedral compilers (blue bars). We use the default Polybench flags, the \texttt{Large} dataset size, and select the best performance of competing compilers from the flags specified in the Appendix~C. On the CPU, we see that for most kernels, the performance of unoptimized SDFGs is closer to that of the polyhedral compilers than to the general-purpose compilers. The cases where SDFGs are on the order of standard compilers are solvers (e.g., \texttt{cholesky}, \texttt{lu}) and simple linear algebra (e.g., \texttt{gemm}). In both cases, data-centric transformations are necessary to optimize the computations, which we exclude from this test in favor of demonstrating SDFG expressibility.

On the GPU, in most cases SDFGs generate code that outperforms PPCG, a tool specifically designed to transform polyhedral applications to GPUs. In particular, the \texttt{bicg} GPU SDFG is 11.8$\times$ faster than PPCG. We attribute these speedups to avoiding unnecessary array copies due to explicit data dependencies, as well as the inherent parallel construction of the data-centric representation.

\vspace{-0.5em}\section{Performance Evaluation}
\label{sec:performance}
\label{sec:perf}

We evaluate the performance of SDFGs on a set of fundamental kernels, followed by three case studies: analysis of matrix multiplication, a graph algorithm, and a real-world physics application.

\paragraph{Experimental Setup}
We run all of our experiments on a server that contains an Intel 12-core Xeon E5-2650 v4 CPU (clocked at 2.20GHz, HyperThreading disabled, 64 GB DDR4 RAM) and a Tesla P100 GPU (16 GB HBM2 RAM) connected over PCI Express. For FPGA results, we use a Xilinx VCU1525 board, hosting an XCVU9P FPGA and four DDR4 banks at 2400 MT/s. We run the experiments 30 times and report the median running time (including memory copy), where the error-bars indicate 95\% confidence interval of all runs, and points are outliers (Tukey fences, 1.5 IQR). For Polybench running times, we use the provided measurement tool, which reports the average time of five runs. All reported results were executed in hardware, including the FPGA.

\paragraph{Compilation}
Generated code from SDFGs is compiled using \texttt{gcc} 8 for CPU results, CUDA 9.2 for GPU, and Xilinx SDAccel 2018.2 for FPGAs. Compilation flags: \texttt{-std=c++14 -O3 -march=native -ffast-math} for CPU, \texttt{-std=c++14 -arch sm\_60 -O3} for GPU, and \texttt{-std=c++11 -O3} for FPGA. Fundamental kernels use single-precision floating point types, Polybench uses the default experiment data-types (mostly double-precision), and graphs use integers.

\vspace{-0.5em}\subsection{Fundamental Computational Kernels}
\label{sec:perf:kernels}

We begin by evaluating 5 core scientific computing kernels, implemented over SDFGs:
\begin{itemize}
	\item \textbf{Matrix Multiplication (MM)} of two 2,048$\times$2,048 matrices.
	\item \textbf{Jacobi Stencil}: A 5-point stencil repeatedly computed on a 2,048 square 2D domain for $T$=1,024 iterations, with constant (zero) boundary conditions.
	\item \textbf{Histogram}: Outputs the number of times each value (evenly binned) occurs in a 8,192 square 2D image.
	\item \textbf{Query}: Filters a column of 67,108,864 elements according to a given condition (filters roughly 50\%).
	\item \textbf{Sparse Matrix-Vector Multiplication (SpMV)} of a CSR matrix (8,192 square; 33,554,432 non-zero values).
\end{itemize}

\begin{figure}[t]
	\begin{subfigure}{\linewidth}
		\centering
		\includegraphics[width=\linewidth]{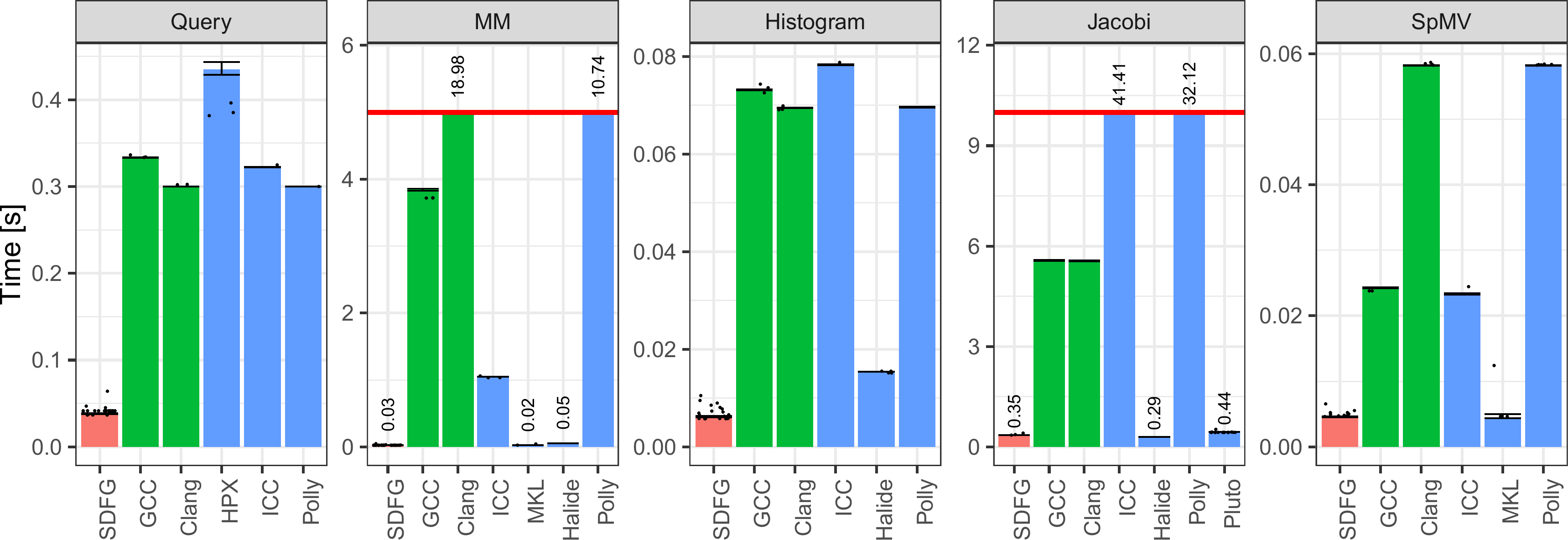}
		\vspace{-1.5em}\caption{CPU}\vspace{1em}
	\end{subfigure}
	\begin{subfigure}{\linewidth}
		\centering
		\includegraphics[width=\linewidth]{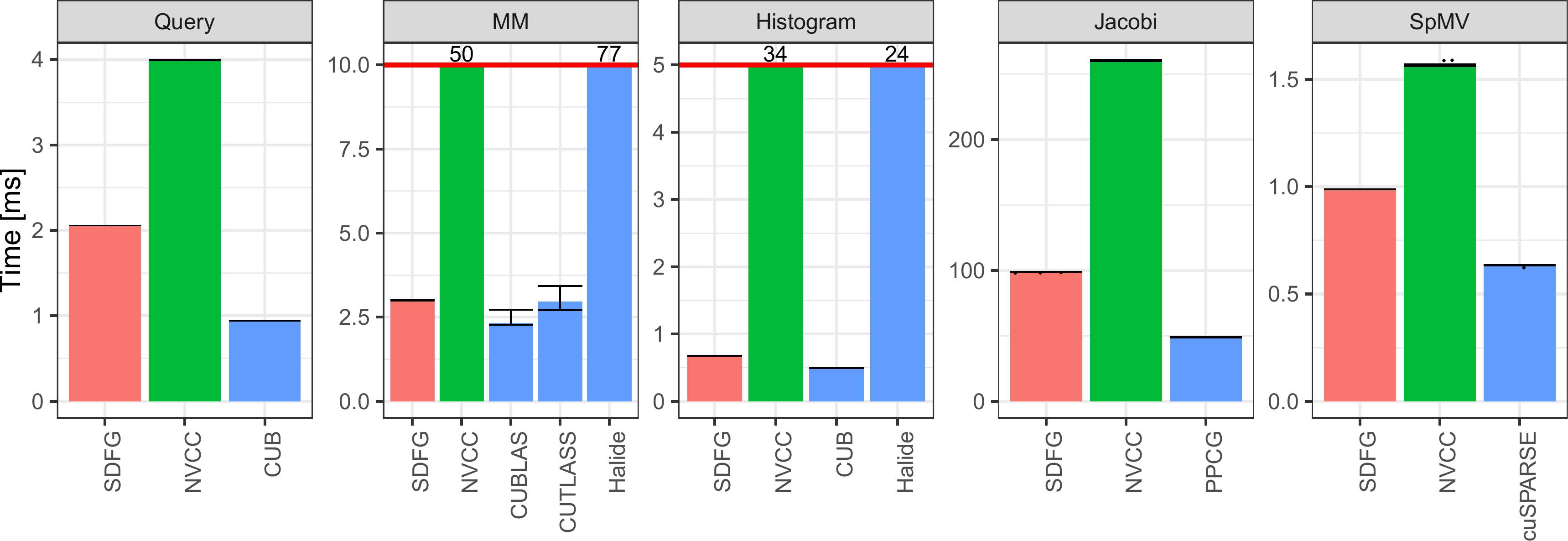}
		\vspace{-2em}\caption{GPU}\vspace{1em}
	\end{subfigure}
	\begin{subfigure}{\linewidth}
		\centering
		\includegraphics[width=\linewidth]{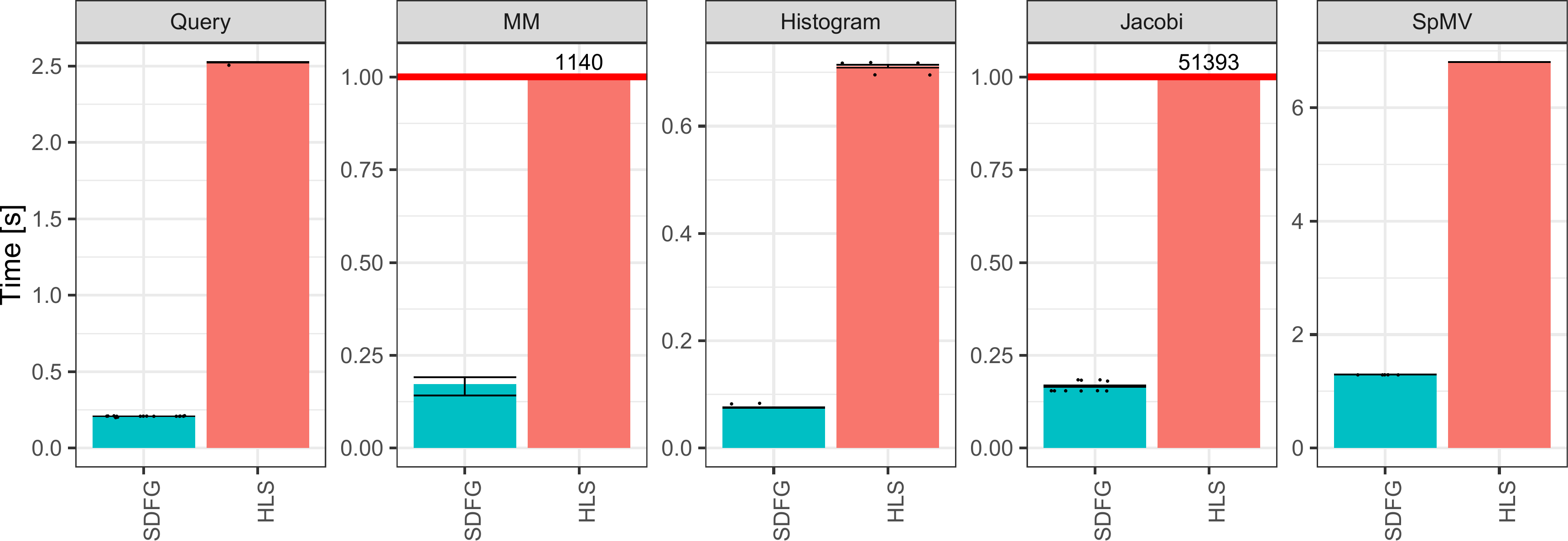}
		\vspace{-1.5em}\caption{FPGA}\vspace{1em}
	\end{subfigure}
	\vspace{-2em}
	\caption{Fundamental Kernel Runtime (lower is better)}
	\label{fig:perf}
\end{figure}

Our results, shown in Fig.~\ref{fig:perf}, are compared with na\"{i}ve implementations of the code, compiled with GCC, Clang, NVCC, and ICC; Intel MKL \cite{mkl} and HPX \cite{hpx} corresponding library calls for CPU; NVIDIA CUBLAS \cite{cublas}, CUSPARSE \cite{cusparse}, and CUB \cite{cub}, as well as Hybrid Hexagonal Tiling over PPCG \cite{ppcg} for GPU; Halide \cite{halide} (auto- and manually-tuned) for CPU and GPU; and Xilinx Vivado~HLS/SDAccel \cite{vivadohls, sdaccel} and Spatial~\cite{spatial} for FPGA.

On all applications, our SDFG results only employ data-centric transformations, keeping tasklets intact (\S \ref{sec:transform}). We highlight key results for all platforms below.

In MM, a highly tuned kernel, SDFGs achieve $\sim$98.6\% of the performance of
MKL, $\sim$70\% of CUBLAS, and $\sim$90\% of CUTLASS, which is the upper bound
of CUDA-based implementations of MM. On FPGA, SDFGs yield a result
4,992$\times$ faster than na\"{i}ve HLS over SDAccel. We also run the
FPGA kernel for $1024\times1024$ matrices and compare to the runtime of
$878\;\text{ms}$ reported for Spatial~\cite{spatial} on the same chip. We
measure $45\;\text{ms}$, yielding a speedup of $19.5\times$.

We observe similar results in SpMV, which is more complicated to optimize due to its irregular memory access characteristics. SDFGs are on par with MKL (99.9\% performance) on CPU, and are successfully vectorized on GPUs.
%

For Histogram, SDFGs enable vectorizing the program, achieving 8$\times$ the performance of gcc, where other compilers fail due to the kernel's data-dependent accesses. 
We implement a two-stage kernel for the FPGA, where the first stage reads
16 element vectors and scatters them to 16 parallel processing elements
generated with map unrolling (similar to Fig.~\ref{fig:systolicarray}), each
computing a separate histogram. In the second stage, the histograms are merged
on the FPGA before copying back the final result. This yields a $10\times$
speedup in hardware. 

In Query, SDFGs are able to use streaming data access to parallelize the
operation automatically, achieving significantly better results than HPX
and STL. On FPGA we read wide vectors, then use a deep pipeline
to pack the sparse filtered vectors into dense vectors. This scheme yields a $10\times$ speedup, similar to Histogram. 

For Jacobi on CPU, we use a domain-specific transformation (\texttt{DiamondTiling}). We see that it outperforms standard implementations
by up to two orders of magnitude, performing 90$\times$ faster than Polly and marginally outperforming
Pluto, which uses a similar tiling technique. 
In Halide, when all timesteps are hard-coded in the pipeline (which takes $\sim$68 minutes to compile due to the stateless dataflow representation), its auto-tuner yields the best result, which is 20\% faster than SDFGs. For FPGAs, Jacobi is
mapped to a systolic array of processing elements, allowing it to scale up with
FPGA logic to $139~\text{GOp}/\text{s}$.
Overall, the results indicate that data-centric transformations can yield competitive performance across both architectures and memory access patterns.

\vspace{-0.5em}\subsection{Case Study I: Matrix Multiplication}\label{sec:perf:mm}

\begin{figure}[t]
	\includegraphics[width=.75\linewidth]{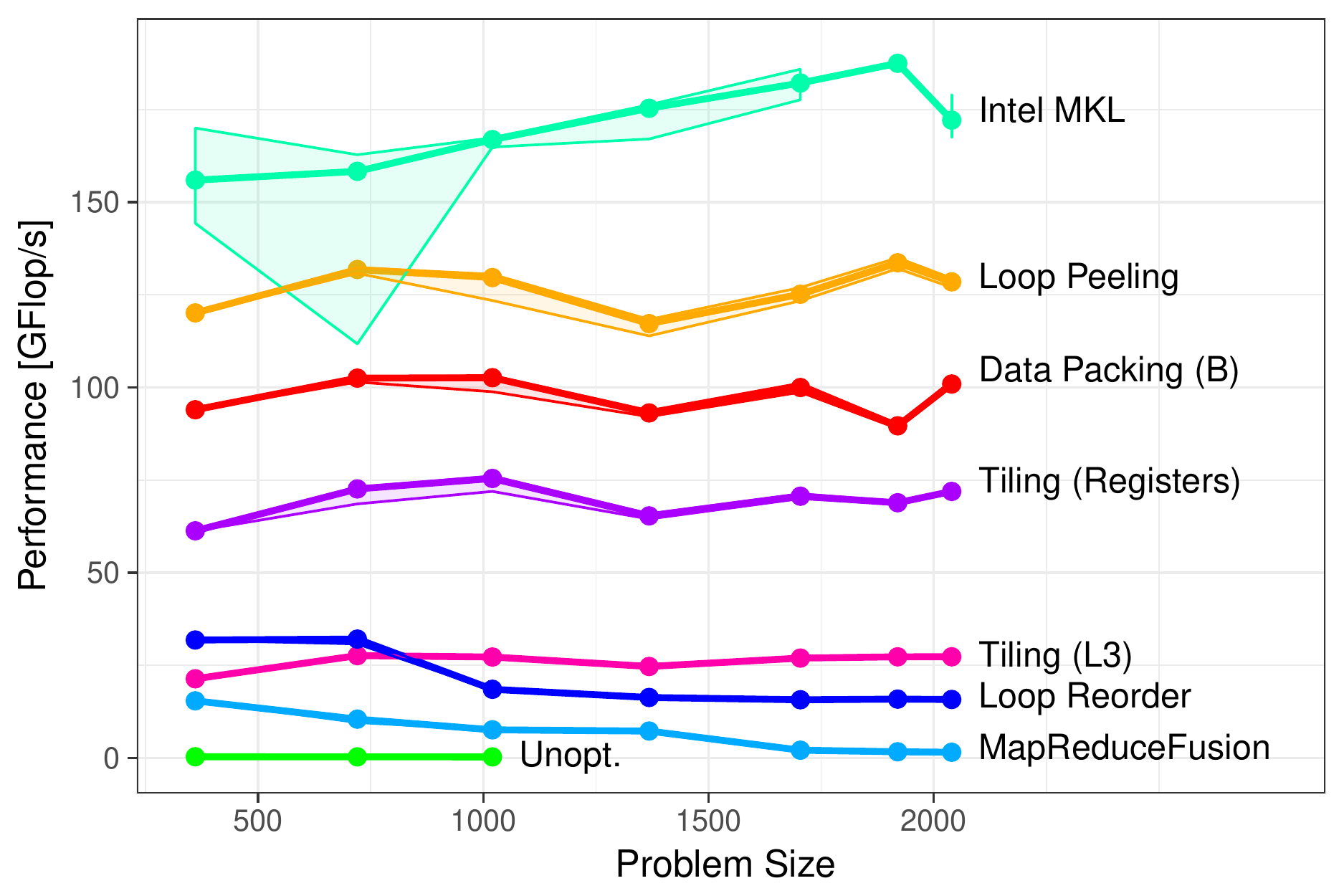}
	\vspace{-1em}
	\caption{Performance of Transformed GEMM SDFG}
	\vspace{-0.5em}
	\label{fig:gemmprog}
\end{figure}

The transformation chain and performance results from Fig.~\ref{fig:sdfg-mm} to the MM CPU SDFG in the previous section are shown in Fig.~\ref{fig:gemmprog}.

After fusing the map and reduction into a write-conflict resolution memlet (\texttt{MapReduceFusion} transformation), we largely follow the approach of Goto and van de Geijn \cite{goto08anatomy}. Manually reordering the map dimensions in the SDFG reduces data movement and yields a marginal improvement. We then tile for the L3 cache (\texttt{MapTiling}) and tile again for mapping to vector registers. To use caches efficiently, we pack the tiles of $B$ and store tiles of $C$ using the \texttt{LocalStorage} transformation (twice), followed by \texttt{Vectorization} to generate code that uses vector extensions. Lastly, we apply a transformation to convert the internal write-conflict resolution memlet into a state machine, peeling the loop (\texttt{ReducePeeling}).

The figure shows that not all transformations yield immediate speedups, yet they are necessary to expose the next steps. Moreover, after only 7 steps the performance increases by $\sim$536x (75\% of MKL), and further increases to 98.6\% of MKL after tuning transformation parameters for a specific size (Fig.~\ref{fig:perf}).

\vspace{-0.5em}\subsection{Case Study II: Graph Algorithms}
\label{sec:perf:graph}

We implement an irregular computation problem on multi-core CPUs: Breadth-First Search (BFS). 
We use the data-driven push-based algorithm \cite{besta17push}, and test five graphs with different characteristics as shown in Appendix~E.

\begin{figure}[t]
	\includegraphics[width=.7\linewidth]{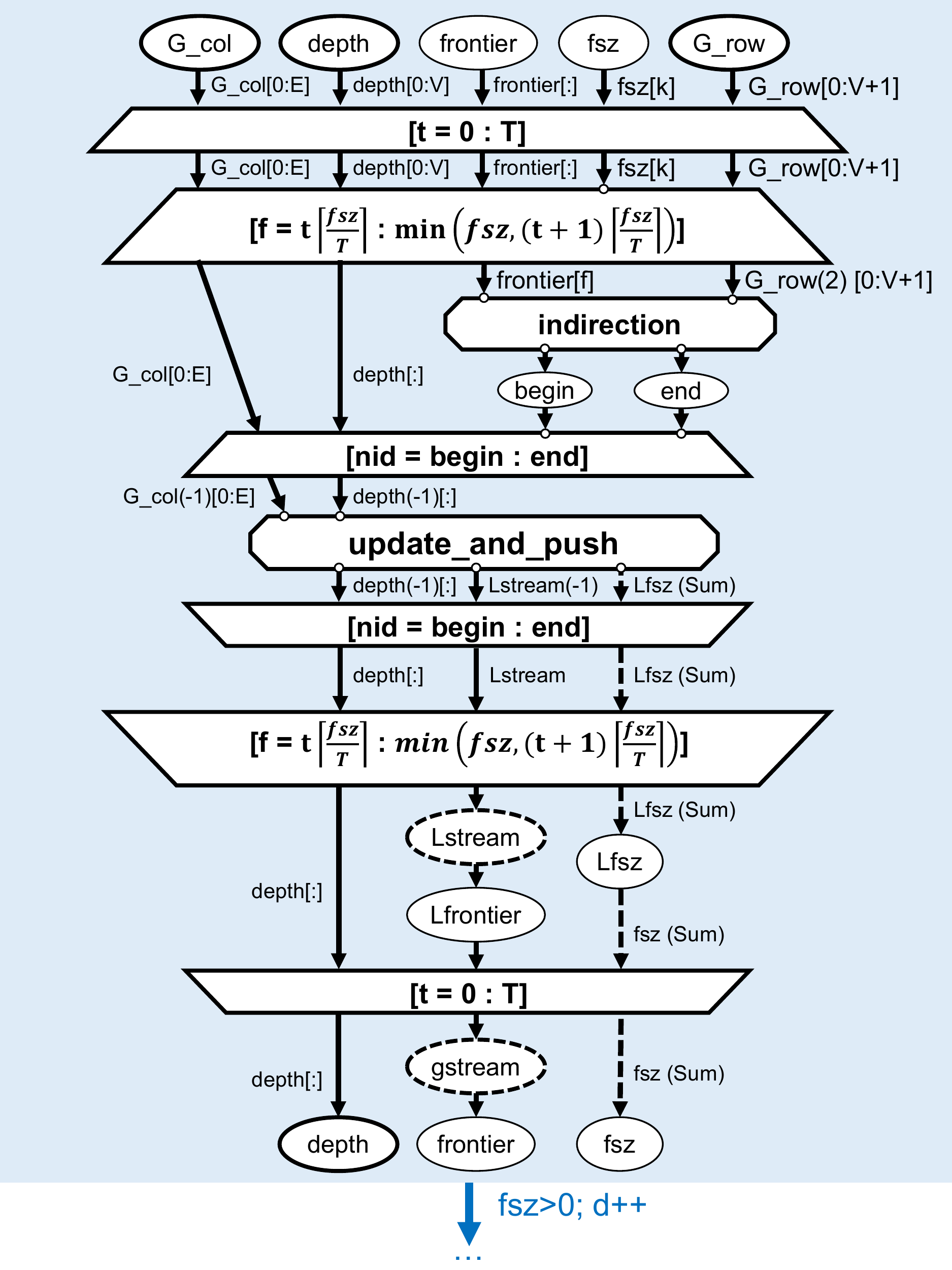}
	\vspace{-1em}
	\caption{Main State of Optimized BFS SDFG}
	\label{fig:bfs}
\end{figure}

Due to the irregular nature of the algorithm, BFS is not a trivial problem to optimize. However, SDFGs inherently support constructing the algorithm using streams and data-dependent map ranges. The primary state of the optimized SDFG is shown in Fig.~\ref{fig:bfs}, which contains only 14 nodes (excluding input/output data) and is the result of \textbf{three} transformations from the base Python program. In particular, the initial generated OpenMP code does not map well to cores due to the dynamic scheduling imposed by the frontier. We mitigate this effect by applying \one{} \texttt{MapTiling} with $T$ tiles. Since the accesses to the next frontier are performed concurrently through \texttt{gstream}, major overhead was incurred due to atomic operations. Using \two{} \texttt{LocalStream}, the results are accumulated to \texttt{Lfrontier} and bulk updates are performed to the global frontier. Similarly, we use \three{} \texttt{LocalStorage} for frontier size accumulation.

\begin{figure}[h!]
	\vspace{-0.75em}
	\includegraphics[width=.7\linewidth]{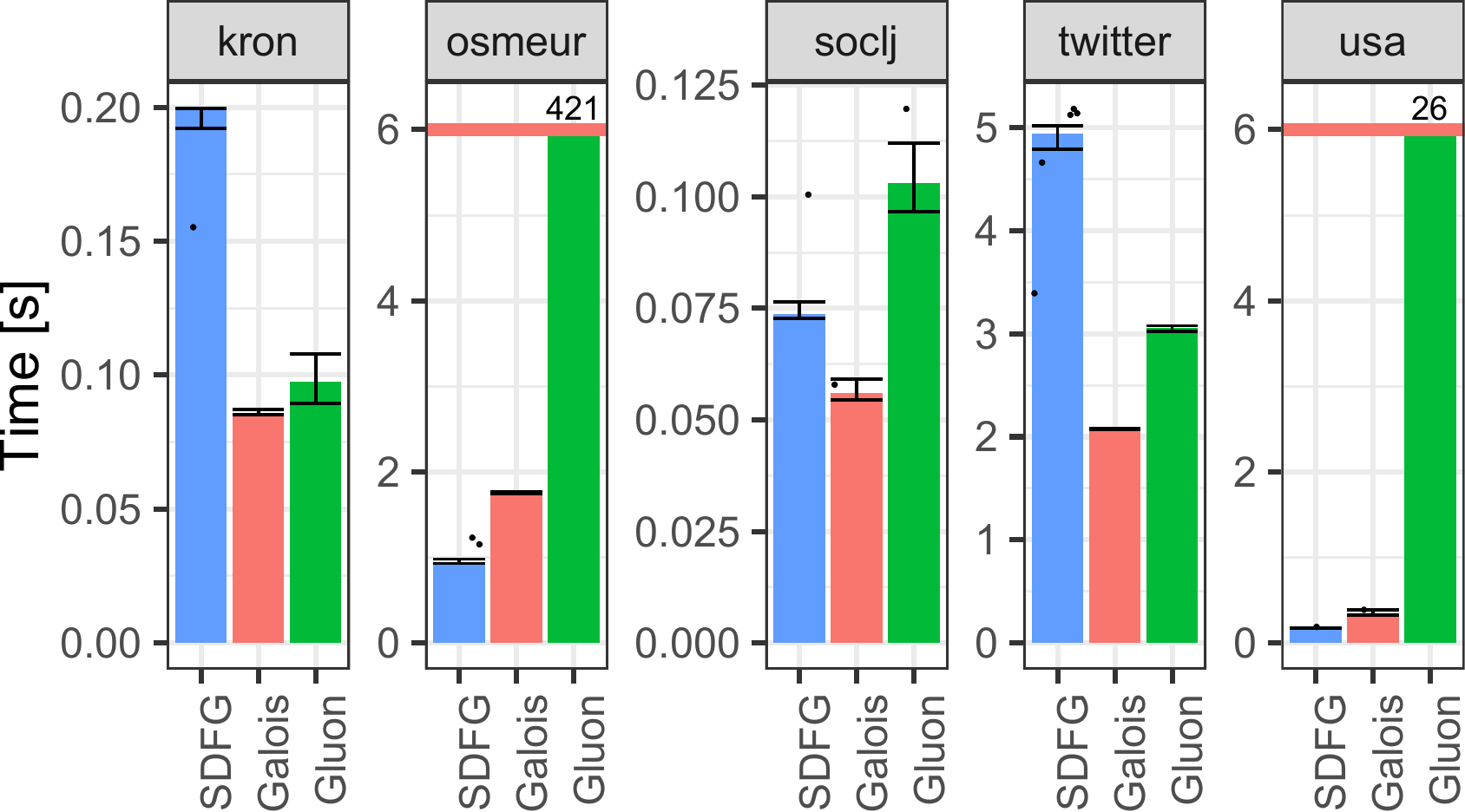}
	\vspace{-1em}
	\caption{BFS Performance}
	\vspace{-1.5em}
	\label{fig:bfsperf}
\end{figure}
\begin{figure*}[t]
	\includegraphics[width=\linewidth]{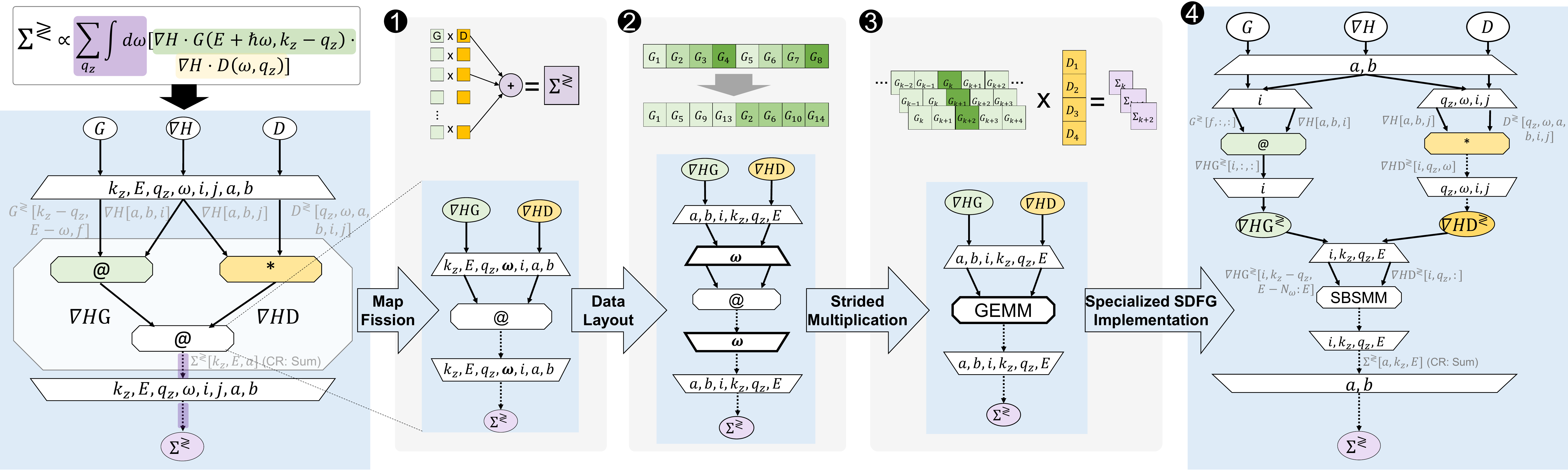}
	\vspace{-1.5em}
	\caption{Optimizing Scattering Self-Energies with SDFG Transformations~\cite{omengb}}
	\vspace{-1em}
	\label{fig:omen}
\end{figure*}

We compare our results with two state-of-the-art CPU graph processing frameworks: Galois \cite{galois} and Gluon \cite{gluon}. We use the default settings (\texttt{bfs\_push} for Gluon, \texttt{SyncTile} for Galois) and use 12 threads (1 thread per core).
In Fig. \ref{fig:bfsperf}, we see that SDFGs perform on-par with the frameworks on all graphs, where Galois is marginally faster on social networks  ($\sim$1.53$\times$ on \texttt{twitter}) and the Kronecker graph. However, in road maps (\texttt{usa}, \texttt{osm-eur}) SDFGs are up to 2$\times$ faster than Galois. This result could stem from our fine-grained scheduling imposed by the map scopes.

\vspace{-0.5em}\subsection{Case Study III: Quantum Transport}

Quantum Transport (QT) Simulation is used for estimating heat dissipation in nanoscale transistors. OMEN~\cite{omen} is an efficient (two-time Gordon Bell finalist) QT simulator based on a nonlinear solver, written in C++ and CUDA using MKL, CUBLAS, and CUSPARSE.

We use SDFGs and transformations to optimize the computation of OMEN, the full details of which are described by Ziogas et al.~\cite{omengb}. A significant portion of the OMEN runtime is spent in computing Scattering Self-Energies (SSE), which is given by the formula in Fig.~\ref{fig:omen} (top-left).
Here we focus on the computation of $\Sigma^\gtrless$. Upon generating the SDFG from the formula, we see that there are multiple parallel dimensions that compute small matrix multiplications (denoted as \texttt{@}) and Hadamard products (\texttt{*}), reducing the result with a summation. 

In step \one, we split the parallel computation into several maps, creating transient 5D and 6D arrays. Steps \two{} and \three{} transform the data layout by reorganizing the map dimensions and transient arrays such that the multiplications can be implemented with one ``batched-strided'' optimized CUBLAS GEMM operation. Lastly, step \four{} replaces the CUBLAS call with a \textit{specialized} nested SDFG that performs small-scale batched-strided matrix multiplication (SBSMM), transformed w.r.t. matrix dimensions and cache sizes in hardware.

\begin{table}[h]
	\begin{center}
		\caption{SSE Performance}\vspace{-1em}
		\small
		\label{tbl:omenperf}
		\begin{tabular}{l rrrr} 
			\toprule
			\bf Variant & \bf Tflop & \bf Time [s] & \bf \% Peak & \bf Speedup \\
			\midrule
			OMEN \cite{omen} & 63.6 & 965.45 & 1.3\% & 1$\times$ \\
			Python (numpy) & 63.6 & 30,560.13 & 0.2\% & 0.03$\times$ \\
			DaCe & 31.8 & \textbf{29.93} & \textbf{20.4\%} & \textbf{32.26}$\times$ \\
			\bottomrule
		\end{tabular}
	\end{center}
\end{table}

\begin{table}[h]
	\setlength{\tabcolsep}{4pt}
	\caption{Strided Matrix Multiplication Performance}\vspace{-1em}
	\label{tbl:sbsmm}
	\small
	\begin{tabular}{lrrrrrr}
		\toprule
		& \multicolumn{3}{c}{\textbf{CUBLAS}} & \multicolumn{3}{c}{\textbf{DaCe (SBSMM)}}\\
		\cmidrule(l{2pt}r{2pt}){2-4}\cmidrule(l{2pt}r{2pt}){5-7}
		\textbf{GPU} & Gflop & Time & \% Peak (Useful) & Gflop & Time & \% Peak \\
		\midrule
		P100 & 27.42 & 6.73 ms & 86.6\% (6.1\%) & 1.92 & \textbf{4.03 ms} & 10.1\% \\
		V100 & 27.42 & 4.62 ms & 84.8\% (5.9\%) & 1.92 & \textbf{0.97 ms} & 28.3\% \\
		
		\bottomrule
	\end{tabular}
\end{table}

Table \ref{tbl:omenperf} lists the overall SSE runtime simulating a 4,864 atom nanostructure over OMEN, numpy (using MKL for dense/sparse LA), and DaCe. Using data-centric transformations, the under-utilization resulting from the multitude of small matrix multiplications is mitigated, translating into a \textbf{32.26$\times$ speedup for SDFGs over manually-tuned implementations}, and 1,021$\times$ over Python. 

Breaking down the speedup, the specialized SDFG implementation of SBSMM (Table \ref{tbl:sbsmm}) outperforms CUBLAS by up to 4.76$\times$. 
This demonstrates that performance engineers can use the data-centric view to \textit{easily tune dataflow for specific problems that are not considered by existing library function implementations}.

\vspace{-0.5em}\section{Related Work}

\begin{figure}[t]
	\centering
	\includegraphics[width=.95\linewidth]{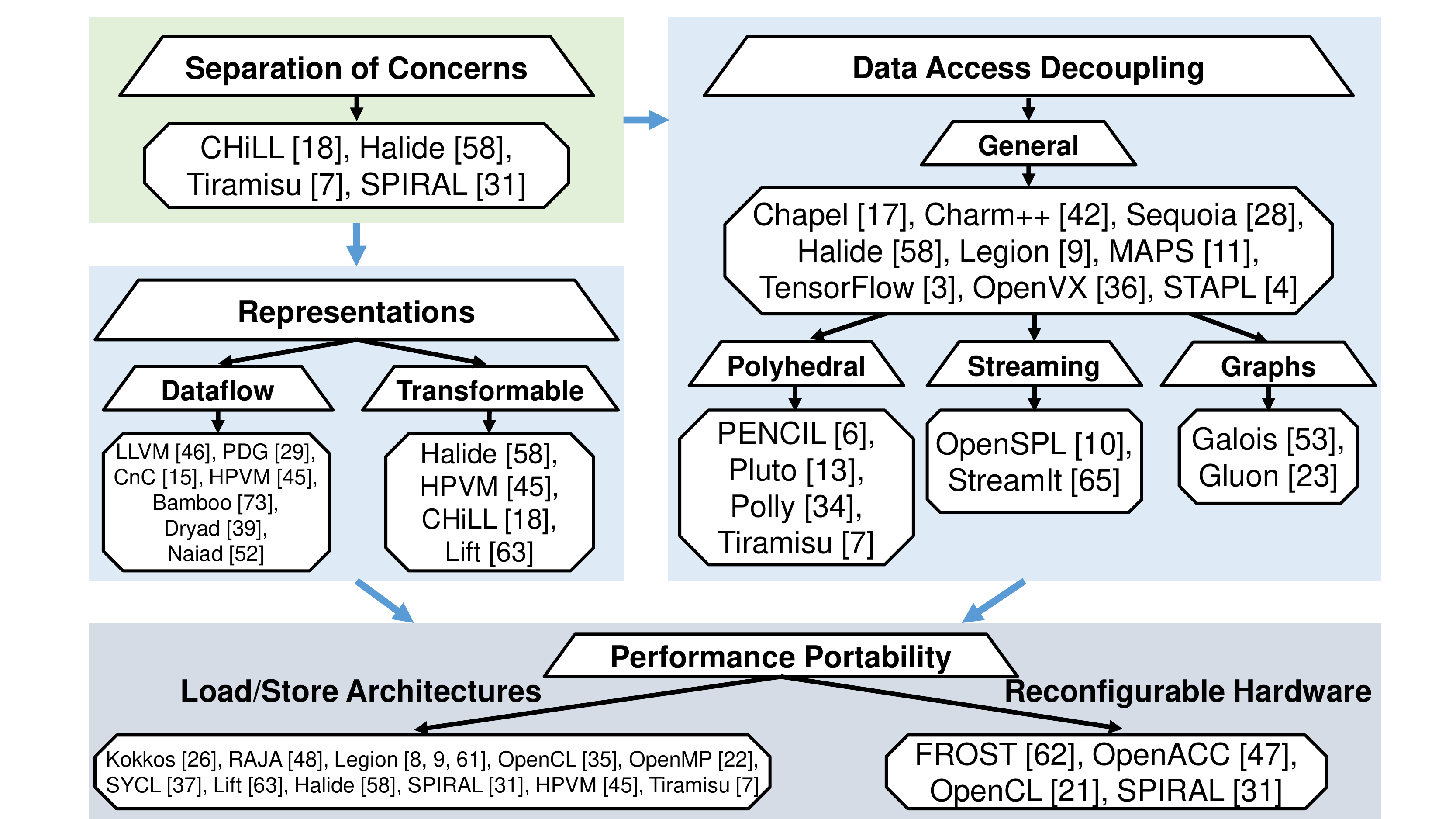}
	\vspace{-1em}
	\caption{Related Work}
	\label{fig:related-work}
\end{figure}

Multiple previous works locally address issues posed in this paper. We discuss those papers below, and summarize them in Fig.~\ref{fig:related-work}.

\paragraph{Performance Portability}
Performance-portable programming models consist of high-performance libraries and programming languages. Kokkos~\cite{kokkos} and RAJA~\cite{raja} are task-based HPC libraries that provide parallel primitives (e.g., forall, reduce) and offer execution policies/spaces for CPUs and GPUs. The Legion~\cite{legion} programming model defines hierarchical parallelism and asynchronous tasking by inferring data dependencies from access sets, called \textit{logical regions}, which are composed of index spaces and accessed fields. Logical regions are similar to subsets in memlets, but differ by implicitly representing data movement. Legate~\cite{legate} is an implementation of the numpy interface that uses Legion to target distributed GPU systems from Python. Language and directive-based standards such as OpenCL \cite{opencl}, OpenACC, and OpenMP \cite{openmp} also provide portability, where some introduce FPGA support through extensions~\cite{acchls,alterahls}. SYCL~\cite{sycl} is an embedded DSL standard extending OpenCL to enable single-source (C++) heterogeneous computing, basing task scheduling on data dependencies. However, true performance portability cannot be achieved with these standards, as optimized code/directives vastly differ on each platform (especially in the case of FPGAs). Other frameworks mentioned below~\cite{chill,lift,halide,hpvm,spiral,tiramisu,regent} also support imperative and massively parallel architectures (CPUs, GPUs), where Halide and Tiramisu have been extended~\cite{frost} to target FPGA kernels. As opposed to SDFGs, none of the above models were designed to natively support both load/store architectures and reconfigurable hardware.

\paragraph{Separation of Concerns}
Multiple frameworks explicitly separate the computational algorithm from subsequent optimization schemes. In CHiLL \cite{chill}, a user may write high-level transformation scripts for existing code, describing sequences of loop transformations. These scripts are then combined with C/C++, FORTRAN or CUDA \cite{cuda-chill} programs to produce 
optimized code using the polyhedral model. Image processing pipelines written in the Halide \cite{halide} embedded DSL are defined as operations, whose schedule is separately generated in the code by invoking commands such as \texttt{tile}, \texttt{vectorize}, and \texttt{parallel}. Tiramisu \cite{tiramisu} operates in a similar manner, enabling loop and data manipulation. In SPIRAL \cite{spiral}, high-level specifications of computational kernels are written in a DSL, followed by using breakdown and rewrite rules to lower them to optimized algorithmic implementations.
SDFGs, along with DIODE and the data-centric transformation workflow, streamlines such approaches and promotes a solution that enables knowledge transfer of optimization techniques across applications.

\paragraph{Dataflow Representations}
Several IRs combine dataflow with control flow in graph form. The LLVM IR \cite{llvm} is a control-flow graph composed of basic blocks of statements. Each block is given in Single Static Assignment (SSA) form and can be transformed into a dataflow DAG. Program Dependence Graphs (PDGs) \cite{ferrante87pdg} represent statements and predicate expressions with nodes, and the edges represent both data dependencies and control flow conditions for execution.
The PDG is a control-centric representation in which statements are assumed to execute dynamically, a model that fits instruction-fetching architectures well. SDFGs, on the other hand, define explicit state machines of dataflow execution, which translate natively to reconfigurable hardware.
Additionally, in PDGs data-dependent access (e.g., resulting from an indirection) creates the same edge as an access of the full array. This inhibits certain transformations that rely on such accesses, and does not enable inferring the total data movement volume, as opposed to the memlet definition.
SDFGs can be trivially converted to the SSA and PDG representations. In the other direction, however, the parametric concurrency that the SDFG scopes offer would be lost (barring specific application classes, e.g., polyhedral).
As the representations operate on a low level, with statements as the unit element, they do not encapsulate the multi-level view as in SDFGs.
HPVM \cite{hpvm} extends the LLVM IR by introducing hierarchical dataflow graphs for mapping to accelerators, yet still lacks a high-level view and explicit state machines that SDFGs offer.
Other representations include Bamboo \cite{bamboo}, an object-oriented dataflow model that tracks state locally through data structure mutation over the course of the program; Dryad \cite{dryad} and Naiad \cite{naiad}, parametric graphs intended for coarse-grained distributed data-parallel applications, where Naiad extends Dryad with definition of loops in a nested context; simplified data dependency graphs for optimization of GPU applications~\cite{wahib14}; deterministic producer/consumer graphs~\cite{cnc}; and other
combinations of task DAGs with data movement~\cite{gajinov12}. As the SDFG provides general-purpose state machines with dataflow, all the above models can be fully represented within it, where SDFGs have the added benefit of encapsulating fine-grained data dependencies.

\paragraph{Data-Centric Transformations}
Several representations \cite{chill,lift,halide,pipes,polymage,sbirlea15} provide a fixed set of high-level program transformations, similar to those presented on SDFGs. In particular, Halide's schedules are by definition data-centric, and the same applies to polyhedral loop transformations in CHiLL. HPVM also applies a number of optimization passes on a higher level than LLVM, such as tiling, node fusion, and mapping of data to GPU constant memory. Lift \cite{lift} programs are written in a high-level functional language with predefined primitives (e.g., map, reduce, split), while a set of rewrite rules is used to optimize the expressions and map them to OpenCL constructs.
Loop transformations and the other aforementioned coarse-grained optimizations are all contained within our class of data-centric graph-based transformations, which can express arbitrary data movement patterns.

\paragraph{Decoupling Data Access and Computation}
The Chapel \cite{chapel} language supports controlling and reasoning about locality by defining object locations and custom iteration spaces. Charm++~\cite{kale93charm} is a parallel programming framework, organized around message passing between collections of objects that perform tasks in shared- or distributed-memory environments.
In the Sequoia \cite{sequoia} programming model, all communication is made explicit by having tasks exchange data through argument passing and calling subtasks.  MAPS \cite{bnsc15} separates data accesses from computation by coupling data with their memory access patterns. This category also includes all frameworks that implement the polyhedral model, including CHiLL, PENCIL \cite{pencil}, Pluto \cite{pluto}, Polly \cite{polly} and Tiramisu. Furthermore, the concept of decoupling data access can be found in models for graph analytics \cite{galois, gluon}, stream processing \cite{openspl, streamit}, machine learning \cite{abadi16}, as well as other high-throughput \cite{openvx} and high-performance \cite{stapl} libraries. Such models enable automatic optimization by reasoning about accessed regions. However, it is assumed that the middleware will carry most of the burden of optimization, and thus frameworks are tuned for existing memory hierarchies and architectures. This does not suffice for fine-tuning kernels to approach peak performance, nor is it portable to new architectures. In these cases, a performance engineer typically resorts to a full re-implementation of the algorithm, as opposed to the workflow proposed here, where SDFG transformations can be customized or extended.

\paragraph{Summary} In essence, SDFGs provide the expressiveness of a general-purpose programming language, while enabling performance portability without interfering with the original code.
Differing from previous models, the SDFG is not limited to specific application classes or hardware, and the extensible data-centric transformations generalize existing code optimization approaches.

\vspace{-0.5em}
\section{Conclusion}
In this paper, we present a novel data-centric model for producing high-performance computing applications from scientific code. Leveraging dataflow tracking and graph rewriting, we enable the role of a performance engineer --- a developer who is well-versed in program optimization, but does not need to comprehend the underlying domain-specific mathematics. We show that by performing transformations on an SDFG alone, i.e., without modifying the input code, it is possible to achieve performance comparable to the state-of-the-art on three fundamentally different platforms.

The IR proposed in this paper can be extended in several directions. Given a collection of transformations, research may be conducted into their systematic application, enabling automatic optimization with reduced human intervention. More information about tasklets, such as arithmetic intensity, can be recovered and added to such automated cost models to augment dataflow captured by memlets. Another direction is the application of SDFGs to distributed systems, where data movement minimization is akin to communication-avoiding formulation.

\begin{acks}
We thank Hussein Harake, Colin McMurtrie, and the whole CSCS team granting
access to the Greina and Daint machines, and for their excellent technical
support.
This project has received funding from the European Research Council (ERC) under the European Union's Horizon 2020 programme (grant agreement DAPP, No. 678880).
T.B.N. is supported by the ETH Zurich Postdoctoral Fellowship and Marie Curie Actions for People COFUND program. 
\end{acks}

\bibliography{references}

\clearpage

\appendix

\section{SDFG Operational Semantics}
\label{app:sec:semantics}

We begin by defining the different elements of our IR, which is a graph
of graphs.  We follow by defining how a function expressed in our IR must be
called and then give semantic rules of how an SDFG is
evaluated. We precede each operational semantic rule by a less formal
description of the intent of the rule.

\subsection{Elements of an SDFG}

An SDFG is a directed multigraph defined by the tuple $(S,$ $T, s_0)$, whose vertex set $S$
represent \emph{states}, its edges $T$ represent \emph{transitions} between
them, and has one \emph{start-state} $s_0$. It is a multigraph since there can be multiple transitions between a pair of states. 

An SDFG state $s\in S$ is a named, directed acyclic multigraph defined by the tuple $(V, E, name)$.
Each node $v_i \in V$ is of one of the
following types, as shown in Table 1 of the paper: \emph{data}, \emph{tasklet}, \emph{mapentry}, \emph{mapexit}, \emph{stream},
\emph{reduce},  \emph{consume-entry},
\emph{consume-exit}, \emph{invoke}. Each node type defines \emph{connectors} --- attachment points for edges defining the nature of the connection, and the edges $E$ indicate dependencies which constrain the execution order.
Each edge carries a \emph{memlet}, an annotation that specifies dataflow, as well as the connectors on its source and destination nodes. Below, we describe the node types in SDFG states and the exact definition of memlets.

As opposed to inherently sequential representations (cf. \cite[Rule 3.15]{impsemantics}), in SDFGs the execution order
is mainly constrained by explicit dependencies. 

To parameterize some aspects of the graph, there exists a global namespace for
\textbf{symbols}. 
Symbols can be used in all \emph{symbolic expressions} mentioned below, and at
runtime they evaluate to scalar values. 
(\S~\ref{sec:execution}).

A \textbf{data} node represents a typed location in memory. The memory itself
can be allocated outside of the SDFG and passed as a pointer upon invocation,
or allocated at runtime if transient. A data node is a tuple (\emph{name}, \emph{basetype},
\emph{dimensions}, \emph{transient}).  The name is an identifier, the basetype
a basic type (e.g., \texttt{int}, \texttt{float}, \texttt{double}), and dimensions a list of symbolic
integer expressions.  Memory pointed to by differently named data nodes must
not alias. A data node has an implicit connector for each adjacent edge.

A \textbf{tasklet} node represents computation. It is defined by the tuple
\textit{(inputs, outputs, code)}, where inputs and outputs are sets of elements of the form
\textit{(name, basetype, dimensions)} that define connectors with the same name, representing the only external memory the code can read or write. Code is a string, which can be written in Python or languages supported by the target architecture, such as C++. 

A \textbf{stream} node represents an array of typed locations in memory, accessed
using queue semantics, i.e., using \texttt{push} and \texttt{pop} operations. A stream has the input connectors \textit{data} and \textit{push}, as well as the output
connectors \textit{data} and \textit{pop}. The \textit{data} connectors allow
initialization/copying of the data stored in the queues from/to an array,
\textit{push} and \textit{pop} enqueue and dequeue elements, respectively.

\begin{figure}[h]
	\includegraphics[width=.6\linewidth,page=4]{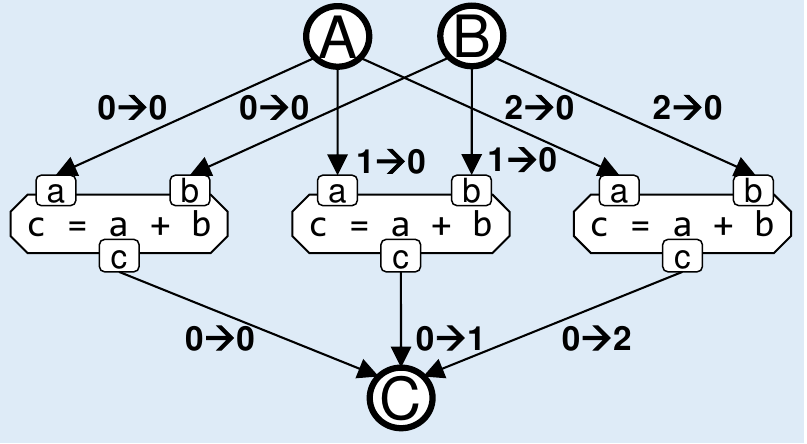}
	\caption{Expanded Map with \textit{subset}$\rightarrow$\textit{reindex} Notation}
	\label{fig:reindex}
\end{figure}

The \textbf{memlet} annotation represents dataflow,
defined by the tuple \textit{(src\_conn, dst\_conn, subset, reindex, accesses, wcr)}. 
The \textit{subset} function selects which
subset of elements visible at the source connector will flow to the destination
connector. The \textit{reindex} function specifies at which indices the data will
be visible at the destination node. We express \textit{subset} and \textit{reindex} as functions on integer sets, and implement them as lists of exclusive ranges, where each range refers to one data dimension and defined by \textit{start:end:stride:tilesize}. 
\textit{tilesize} is used to propagate multiple elements at a time, e.g., vector types. Fig.~\ref{fig:reindex} extends Fig.~5b from the paper by showing the same memlets in subset/reindex notation rather than array index labels.

The \textbf{wcr} optional attribute (write-conflict resolution) is a function $S
\times S \rightarrow S$ for a data type $S$.
The function receives the old value currently present at the destination and the new
value, and produces the value that will be written. 
This attribute is normally defined when data flows concurrently from multiple sources, e.g., in a parallel dot product.

\textbf{Map entry/exit} nodes are defined by a tuple \textit{(range, inputs, outputs)}, creating a scope.
The subgraph within the scope of a map is defined by the nodes that are dominated by the map-entry and post-dominated by the map-exit. This subgraph is expanded at runtime and executed concurrently according to the symbolic \textit{range} of the mapentry, which takes the form \textit{identifier=begin:end:step}. Input and output connectors of map nodes are either defined as \texttt{IN\_*}/\texttt{OUT\_*} (\texttt{*} is any string) to map memlets outside the scope to memlets within it; or other identifiers, which can be used to create data-dependent map ranges.
The range identifier becomes a scope-wide symbol, and can used to define memlet attributes (\textit{subset}, \textit{reindex}).

The \textbf{invoke} node allows to call an SDFG within an SDFG state. Its
semantics are equivalent to a tasklet that has input connectors for each
data-node and undefined symbol used in the invoked SDFG, and an output node for
each data node. 

The \textbf{reduce} alias node is used to implement target-optimized reduction procedures. It is defined by axes and a \textit{wcr} function, consisting of an input and an output connector (of the same type).
It is semantically equivalent to a map whose range spans the incoming and outgoing memlet data, containing an identity tasklet (\texttt{output=input}) and an output memlet annotated by the given \textit{wcr} for the reduced axes.

\textbf{Consume} entry/exit nodes complement maps, enabling producer/consumer relationships via dynamic parallel processing of streams. The two nodes form a scope, similarly to a map, but also accept a \textit{stream\_in} connector (connected to a stream), and a quiescence condition after which processing stops. Thus, a consume entry node is a tuple \textit{(range, cond, inputs, outputs)}. Semantically, a consume scope is equivalent to a map using the same $range$, with an internal \textit{invoke}d SDFG containing the scope. This state connects back to itself without any assignments and the condition $cond$.

A \textbf{state transition} is defined as a tuple \textit{(source, destination,
	condition, assignments)}. \textit{Condition} can depend on 
symbols and data from data nodes in the source state, whereas \textit{assignments} take the form \textit{symbol = expression}. Once a state finishes execution, all outgoing state transitions of
that state are evaluated in an arbitrary order, and the destination of the first
transition whose condition is true is the next state which will be executed. 
If no transition evaluates to true, the program terminates.
Before starting the execution of the next state, all assignments are performed,
and the left-hand side of assignments become symbols.

\subsection{Operational Semantics}
\label{sec:execution}

\subsubsection{Initialization}

\paragraph{Notation}
We denote collections (sets/lists) as capital
letters and their members with the corresponding lowercase letter and a
subscript, i.e., in an SDFG $G=(S, T, s_0)$ the set of states $S$ as $s_i$, with $0 \leq i < |S|$.
Without loss of generality we assume $s_0$ to be the start state.  We denote
the value stored at memory location $a$ as $M[a]$, and assume all basic types
are size-one elements to simplify address calculations.

The state of execution is denoted by $\rho$. Within the state we carry several
sets: $loc$, which maps names of data nodes and transients to memory addresses;
$sym$, which maps symbol names (identifiers) to their current value;
and $vis$, which maps connectors to the data visible at that connector in the current state of execution.

We define a helper function $size()$, which returns the product of all
dimensions of the data node or element given as argument (using $\rho$ to resolve symbolic values).
Furthermore, $id()$ returns the \textit{name} property of a data or transient node, and \textit{offs()} the offset of a data element relative to the start of the memory region it is stored in.
The function $copy()$ creates a copy of the object given as argument, i.e., when we modify the copy, the original object remains the same.

\paragraph{Invocation}
When an SDFG $G$ is called with the data arguments $A \equiv [a_i = p_i]$ ($a_i$ is an identifier, $p_i$ is
an address/pointer) and symbol arguments $Z \equiv [z_i = v_i]$ ($z_i$ is an
identifier, $v_i$ an integer) we initialize the configuration $\rho$:
\begin{enumerate}
	\item For all symbols $z_i$ in $Z$: $sym[z_i] \leftarrow v_i$.
	\item For all data and stream nodes $d_i \in G$ without incoming edges s.t. $id(d_i) = a_i$: $loc(d_i) \leftarrow p_i, vis(d_i.data) \leftarrow M[p_i, ..., p_i + size(d_i)]$.
	\item Set $current$ to a copy of the start state  of $G$, $s_0$.
	\item Set $state$ to $id(s_0)$.
	\item Set $qsize[f_i]$ to zero for all stream nodes $f_i \in G$.
	
\end{enumerate}
This can be expressed as the following rule:

\vspace{-0.5em}
\tiny
\[
\frac{
	\begin{array}{c} G=(S,T), \\
	\mathit{start\_state}(G) \rightarrow s_0, \\
	D: \text{data nodes in } s_0, \\
	F: \text{stream nodes in } s_0 \\
	\end{array}
}
{
	\begin{array}{ll}
	(\mathit{call}(G, A, Z), \rho) \rightarrow \rho[ & \mathit{state} \mapsto \mathit{id}(s_0), \\ 
	& \mathit{current} \mapsto \mathit{copy}(s_0), \\
	& \forall a_i=p_i \in A: \mathit{loc}[a_i \mapsto p_i], \\
	& \forall z_i=v_i \in Z: \mathit{sym}[z_i] \mapsto v_i, \\
	& \forall d_i \in D: \mathit{vis}[d_i.\mathit{data}] \mapsto M[p_i, .., p_i+size(d_i)], \\
	& \forall f_i \in G.S: \mathit{qsize}[\mathit{id}(f_i)] \mapsto 0]
	
	\end{array}
}
\]
\normalsize

\subsubsection{Propagating Data in a State}
\label{sec:dataprop}

Execution of a state entails propagating data along edges, governed by the rules defined below.

Data is propagated within a state in an arbitrary order, constrained only by
dependencies expressed by edges. We begin by considering all nodes in the current state, which we add to \textit{current}. We gradually remove nodes as they are processed, until \textit{current} is empty and we can proceed to the next state or termination.

To keep the description concise, we omit details of address translation and indexing. We also use the \textit{subset} and \textit{reindex} properties as functions that resolve symbols and return data elements directly.

\paragraph{Element Processing}
In each step, we take one element $q$ (either a memlet or a node) of $\mathit{current}$,
for which all input connectors have visible data, then:\\
If $q$ is a \textbf{memlet} \textit{(src, dst, subset, reindex, wcr)}, update $\mathit{vis}[\mathit{dst}]$ to $\mathit{wcr(reindex(subset(vis[src])))}$:

\tiny
\[
\frac{
	\begin{array}{c}
	q = \mathit{memlet}(\mathit{src}, \mathit{dst}, \mathit{subset}, \mathit{reindex}, \mathit{wcr}), \\
	(\mathit{vis}[\mathit{src}], \rho) \neq \emptyset, \\
	(\mathit{wcr(reindex(subset(vis[src])))}, \rho) \rightarrow [d_0, .., d_n]
	\end{array}
}{
	(q, \rho) \rightarrow \rho[ \mathit{vis}[\mathit{dst}] \mapsto [d_0, .., d_n] ]
}
\]
\normalsize

\noindent If $q$ is a \textbf{data} node, update its referenced memory for an input connector $c_i$,\\ $M[\mathit{loc}(\mathit{id}(q)), .., \mathit{loc}(\mathit{id}(q))+\mathit{size}(\mathit{vis}[\mathit{q.data}])] = \mathit{vis}[\mathit{q.data}]$:

\tiny
\[
\frac{
	\begin{array}{c}
	q = \mathit{data(id, dims, bt, \mathit{transient})}, \\
	(\forall x,q.c_i \in \mathit{current}: \mathit{vis}[q.c_i], \rho) \neq \emptyset, \\
	(\forall x,q.c_i \in \mathit{current}: \mathit{vis}[q.c_i], \rho) \rightarrow [d^{i}_{0}, d^{i}_{1}, .., d^{i}_{k_i}], \\
	\forall d^{i}_{j} : loc(id(q)) + \mathit{offs}(d^{i}_{j}) = l^{i}_{j}
	\end{array}
}{
	(q, \rho) \rightarrow \rho [\forall i,k_i: M[l^{i}_{j}, .., l^{i}_{j}+\mathit{size}(d^{i}_{j})] = d^{i}_{j} ]
}
\]
\normalsize

\noindent If $q$ is a \textbf{tasklet}, generate a prologue that allocates local variables for
all input connectors $c_i$ of $q$, initialized to $vis[c_i]$ ($P_1$), as well
as output connectors ($P_2$).  Generate an epilogue $Ep$ which updates
$\rho[vis[c_i] \mapsto v_i ]$ for each output connector $c_i$ of $q$ with the
contents of the appropriate variable (declared in $P_2$). Execute the
concatenation of $(P_1; P_2; code; Ep;)$.  

\tiny 
\[ 
\frac{ 
	\begin{array}{c} 
	q = \mathit{tasklet(Cin, Cout, code)}, \\ 
	(\forall x,q.c_i \in \mathit{current}: vis[q.c_i], \rho) \neq \emptyset, \\ (
	P_1 = [\forall c_i \in Cin: type(c_i) id(c_i) = vis[c_i]];, \rho), \\ 
	P_2 = [\forall c_i \in Cout: type(c_i) id(c_i);], \\
	(Ep = [\forall c_i \in Cout: \&vis(c_i) = id(c_i);], \rho), \\
	\end{array} 
}{ 
	\begin{array}{c}
	(q, \rho) \rightarrow \rho [ exec(P_1; P_2; code; Ep;) ] 
	\end{array} } 
\]
\normalsize

\noindent If $q$ is a \textbf{mapentry} node with range $y = R$ ($y$ is the identifier) and scope $o\subset V$: Remove $o$
from \textit{current}. Remove $q$ and the corresponding map exit node from $o$. For
each element in $r_i \in R$, replicate $o$, resolve any occurrence of $y$ to
$r_i$, connect incoming connectors of $q$ and $p$ in $state$.

\tiny
\[
\frac{
	\begin{array}{l}
	q = \mathit{mapentry(Cin, Cout, R)}, \\
	\forall cin_i : vis[cin_i] \neq \emptyset, \\
	o = \mathit{scope}(q), o' = \mathit{scope}(q) \setminus \{q, \mathit{mexit}(q)\}, \\
	\mathit{NewSyms} = [cin_i : \nexists (cin_i, y)]
	\end{array}
}{
	\begin{array}{ll}
	(q, \rho) \rightarrow \rho[ 
	& \mathit{current} \mapsto o' \cup [\forall r_i : \mathit{ressym}(\mathit{copy}(o'), r_i)], \\
	& \forall n_i \in \mathit{NewSym}: sym[n_i] \mapsto vis[cin_i]~~] 
	\end{array}
}
\]
\normalsize

\noindent If $q$ is a \textbf{consume-entry} node, defined by \textit{(range, cond, cin, cout)}, replace $q$ with a mapentry 
and do the same
for the corresponding consume exit node. Then we create a new SDFG $new$, which contains the contents of the original consume scope $scope(q)$. $new$ consists of one state $s_0$, and a single state transition to the same state with the condition $cond$, defined by $(s_0, s_0, cond, [])$. Finally, we replace $scope(q)$ in \textit{current} with an invoke node for
$new$ and reconnect the appropriate edges between the entry and exit nodes. 

\tiny
\[
\frac{
	\begin{array}{l}
	q = \mathit{consume-entry(range, cond, cin, cout)} \\
	\mathit{newsdfg} = \mathit{SDFG(scope(q) \setminus \{q, cexit(q)\}, (s_0,s_0, cond, []))} \\
	iv = \mathit{invoke}(\mathit{newsdfg}) \\
	men = \mathit{mapentry(range, cin, cout)} \\
	mex = \mathit{mapexit(cexit(q).cin, cexit(q).cout)}
	\end{array}
}{
	(q, \rho) \rightarrow \rho[ \mathit{current} \mapsto (\mathit{current} \setminus {q, cexit(q)}) \cup \{ iv, \mathit{men, mex} \} ]
}
\]
\normalsize

\noindent If $q$ is a \textbf{reduce} node defined by the tuple \textit{(cin, cout, range)}, we create a mapentry
node $men$ with the same range, a mapexit node $mex$, and a tasklet \texttt{o = i}. We add these nodes to the node set of \textit{current}, \textit{nd(current)}. We
connect them by adding edges to the edge set of \textit{current}.

\tiny
\[
\frac{
	\begin{array}{c}
	q = \mathit{reduce(cin, cout, range, wcr)}, \\
	vis[cin] \neq \emptyset, \\
	men = \textit{mapentry}(\{IN\_1\}, \{OUT\_1\}, range), \\
	mex = \textit{mapexit}(\{IN\_1\}, \{OUT\_1\}), \\
	t = \textit{tasklet}(\{i\},\{o\}, ``o=i;")
	\end{array}
}{\arraycolsep=1.4pt
	\begin{array}{ll}
	(q, \rho) \rightarrow \rho [ & nd(current) \mapsto nd(current) \cup \{men, mex, t\} \\
	& ed(current) \mapsto ed(current) \cup \{ (men.OUT\_1, t.i), (t.o, mex.IN\_1) \} \\
	& vis[men.IN\_1] \mapsto vis[q.cin], \\
	& vis[q.IN] \mapsto \emptyset ]
	\end{array}
}
\]
\normalsize

\noindent If $q$ is a \textbf{stream}, we add the data visible at the \textit{push} connector to the
appropriate memory location (indicated by $qsize[id(q)]$), and increment it.
A stream node with an edge connected to its pop connector can only be evaluated if $qsize[id(q)] \geq 1$. When propagating through streams, there are three cases for one step:\\
\one~ Data is both pushed into and popped from the stream:

\tiny
\[
\frac{
	\begin{array}{l}
	q = \mathit{stream(push, pop, size)}, \\
	vis[push] \neq emptyset, \\
	\exists (q.pop, dst) \in current, \\
	\rho[qsize(id(q)) \geq 1]
	\end{array}
}{
	\begin{array}{ll}
	(q, \rho) \rightarrow \rho [ & vis[q.pop] \mapsto M[\mathit{offs}(q), .., \mathit{offs}(q)+size(q)], \\
	& M[\mathit{offs}(q)+size(q) \times qsize[q], .., \\
	& \quad \mathit{offs}(q) + size(q) \times (qsize[q]+1)] \mapsto vis[push],]
	\end{array}
}
\]
\normalsize

\noindent\two~ Data is only pushed to the stream node:

\tiny
\[
\frac{
	\begin{array}{l}
	q = \mathit{stream(push, pop, size)}, \\
	vis[push] \neq emptyset, \\
	\nexists (q.pop, dst) \in current 
	\end{array}
}{
	\begin{array}{ll}
	(q, \rho) \rightarrow \rho [ & M[\mathit{offs}(q)+size(q) qsize[q], .., \\
	& \quad \mathit{offs}(q) + size(q) (qsize[q]+1)] \mapsto vis[push], \\
	& qsize[q] \mapsto qsize[q]+1 ]
	\end{array}
}
\]
\normalsize

\noindent \three~ Data is popped from the stream but not pushed into it:

\tiny
\[
\frac{
	\begin{array}{l}
	q = \mathit{stream(push, pop, size)}, \\
	\nexists (src, q.push) \in S_{state}, \\
	\exists (q.pop, dst) \in current \\
	\end{array}
}{
	\begin{array}{ll}
	(q, \rho) \rightarrow \rho [ & vis[q.pop] \mapsto M[\mathit{offs}(q), .., \mathit{offs}(q)+size(q)], \\ 
	& qsize[q] \mapsto qsize[q]-1 ]
	\end{array}
}
\]
\normalsize

Following the element processing step, we remove $q$ from \textit{current}, repeating the above step until \textit{current} is empty. 

\subsubsection{Evaluating State Transitions}

Once \textit{current} is empty, we evaluate all outgoing state
transitions of the current state:\\$(state, next, cond, assigs) \in T$. For
each transition, we resolve all symbols in $cond$ and the right-hand sides of $assigs$ using $\rho$, then evaluate arithmetic and boolean expressions using standard semantic rules, which we omit here.  If no condition evaluates to true, signal the completion of $G$ to the caller and stop the evaluation of $G$:

\tiny
\[
\frac{
	\begin{array}{c}
	\mathit{current} = \emptyset, \\
	\nexists (\rho[state], next, cond, assigs) \in T: (cond, \rho) \rightarrow True \\
	\end{array}
}{
	\rho[ state \mapsto \emptyset ]
}
\]
\normalsize

Otherwise, we choose an arbitrary transition for which $cond \rightarrow true$ and update $\rho$:
Set $state$ to $next$, set $curr$ to a copy of $S[next]$. For each left-hand side of an assignment
$z_i$, update $sym[z_i]$ with the value of the
corresponding right-hand side $v_i$.  Data propagation then follows Section~\ref{sec:dataprop}:

\tiny
\[
\frac{
	\begin{array}{c}
	$current$ = \emptyset, \\
	\exists (\rho[state], next, cond, assigs) \in T: (cond, \rho) \rightarrow True, \\
	(assigs, \rho) \rightarrow [z_i = v_i]
	\end{array}
}{
	\begin{array}{ll}
	\rho[ & state \mapsto next, \\
	& current \mapsto copy(next), \\
	& sym[z_i] \mapsto v_i ]
	\end{array}
}
\]
\normalsize

\newpage
\section{Data-Centric Graph Transformations}
\label{app:sec:transform}

\begin{table}[t]
	\caption{Supported Transformations}
	\vspace{-1em}
	\label{tbl:trans}
	\centering
	\small
	\begin{tabular}{p{2.2cm} p{5.6cm}}
		\toprule
		\bf Name & \bf Description\\
		\midrule
		\multicolumn{2}{l}{\bf Map transformations}\\\addlinespace
		\texttt{MapCollapse} & Collapses two nested maps into one. The new map has the union of the dimensions of the original maps.\\
		\texttt{MapExpansion} & Expands a multi-dimensional map to two nested ones. The dimensions are split to two disjoint subsets, one for each new map.\\
		\texttt{MapFusion} & Fuses two consecutive maps that have the same dimensions and range.\\
		\texttt{MapInterchange} & Interchanges the position of two nested maps.\\
		\texttt{MapReduceFusion} & Fuses a map and a reduction node with the same dimensions, using conflict resolution.\\        
		\texttt{MapTiling} & Applies orthogonal tiling to a map.\\
		\midrule
		\multicolumn{2}{l}{\bf Data transformations}\\\addlinespace
		\texttt{DoubleBuffering} & Pipelines writing to and processing from a transient using two buffers.\\
		\texttt{LocalStorage} & Introduces a transient for caching data.\\
		\texttt{LocalStream} & Accumulates data to a local transient stream.\\
		\texttt{Vectorization} & Alters the data accesses to use vectors.\\
		\midrule
		\multicolumn{2}{l}{\bf Control-flow transformations}\\\addlinespace
		\texttt{MapToForLoop} & Converts a map to a for-loop.\\        
		\texttt{StateFusion} & Fuses two states into one.\\
		\texttt{InlineSDFG} & Inlines a single-state nested SDFG into a state.\\
		\midrule
		\multicolumn{2}{l}{\bf Hardware mapping transformations}\\\addlinespace
		\texttt{FPGATransform} & Converts a CPU SDFG to be fully invoked on an FPGA, copying memory to the device.\\        
		\texttt{GPUTransform} & Converts a CPU SDFG to run on a GPU, copying memory to it and executing kernels.\\
		\texttt{MPITransform} & Converts a CPU Map to run using MPI, assigning work to ranks.\\
		\bottomrule
	\end{tabular}
	\vspace{1em}
\end{table}

Table \ref{tbl:trans} lists the transformations used in the Performance Evaluation section of the paper.

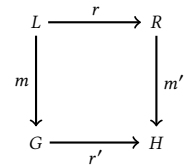
\begin{figure}[h]
	\begin{tikzpicture}[thick,scale=0.8, every node/.style={scale=0.8},shorten >=1pt,->]
	\tikzstyle{vertex}=[]
	
	\node[vertex] (N6) at (0, 0) {$L$};
	\node[vertex] (N7) at (2, 0) {$R$};
	\node[vertex] (N8) at (0, -2) {$G$};
	\node[vertex] (N9) at (2, -2) {$H$};
	
	\draw (N6) -- (N7) node[midway, above] {$r$};
	\draw (N6) -- (N8) node[midway, left] {$m$};
	\draw (N8) -- (N9) node[midway, below] {$r^\prime$};
	\draw (N7) -- (N9) node[midway, right] {$m^\prime$};
	
	\end{tikzpicture}
	\vspace{-1em}
	\caption{Single-Pushout Approach}
	\vspace{-0.5em}
	\label{fig:pushout}
\end{figure}

The implementation of graph transformations is based on
\textit{algebraic graph rewriting} \cite{agrewriting}. Each transformation is
implemented by a \textit{rule} $p: L \xrightarrow{r} R$, which consists of the
two sub-graphs $L$ and $R$, while $r$ is a relation that maps the vertices and
edges of $L$ to elements of the same kind in $R$. Moreover, a specific
\textit{matching} of $L$ in the SDFG $G$ is represented by a relation
$m: L \rightarrow G$. Applying the optimization on $G$ produces the transformed
graph $H$, which can be constructed as part of the \textit{pushout}
$<H, m^\prime, r^\prime>$ of $p$, $G$ and $m$. A visualization of the above
method, also known as the \textit{single-pushout approach}
\cite{spushout}, is shown in Fig. \ref{fig:pushout}.

\section{Polybench Flags}
\label{app:polybench}

We compile Polybench code for all compilers using the following base flags: \texttt{-O3 -march=native -mtune=native}. Each compiler was also tested with different variants of flags, where we report the best-performing result in the paper. The flags are (variants separated by semicolons): 
\begin{itemize}
	\item \textbf{gcc 8.2.0:} \texttt{-O2}; Base flags
	\item \textbf{clang 6.0:} Base flags
	\item \textbf{icc 18.0.3:} Base flags; \texttt{-mkl=parallel -parallel}
	\item \textbf{Polly (over clang 6.0):} \texttt{-mllvm -polly};\\ \texttt{-mllvm -polly -mllvm -polly-parallel -lgomp}
	\item \textbf{Pluto 0.11.4}: \texttt{-ftree-vectorize -fopenmp} and\\ among best of \texttt{--tile}; \texttt{--tile --parallel};\\ \texttt{--tile --parallel --partlbtile};\\ \texttt{--tile --parallel --lbtile --multipar}
	\item \textbf{PPCG 0.8:} Base flags
\end{itemize}

All variants were also tested with compile-time size specialization.

\section{Transformation Example: Redundant Array Removal}

A simple transformation implemented in DaCe is \texttt{RedundantArray}, whose code can be found below.
The transformation removes a transient array that is used directly before another array, creating a copy, and not used anywhere else (making the copy redundant). This situation often happens after transformations and due to the strict nature of some language frontends (e.g., TensorFlow). The subgraph expression to match (lines 6--14) is a path graph of size two, connecting the two access nodes. The nodes have no restriction on their content (see lines 6--7). The function \texttt{can\_be\_applied} is called on a matching subgraph for further programmatic checks (lines 16--58), and the function \texttt{apply} (lines 60--78) applies the transformation using the SDFG builder API. The checks ensure that the array is indeed transient and not used in other instances of data access nodes. To avoid recomputing subsets (which may not be feasible to compute symbolically), if the transformation operates in strict mode, it only matches two arrays of the same shape (lines 51--56). The transformation then operates in a straightforward manner, renaming the memlets to point to the second (not removed) array (lines 66--70) and redirecting dataflow edges to that data access node (lines 73--74). Lastly, the \texttt{in\_array} node is removed from the SDFG state (line 78).

In line 81, the transformation is registered with a global registry. This process enables users to import custom transformations and integrate them into DaCe easily.
However, because \texttt{RedundantArray} is a strict transformation (i.e., can only improve performance), it is also hardcoded in a set of such transformations within the SDFG class. Strict transformations are applied automatically after processing a DaCe program, and other such transformations include \texttt{StateFusion} and \texttt{InlineSDFG}.

\newpage
\begin{lstlisting}
class RedundantArray(pm.Transformation):
    """ Implements the redundant array removal transformation, 
        applied when a transient array is copied to and from (to 
        another array), but never used anywhere else. """

    _in_array = nodes.AccessNode('_')
    _out_array = nodes.AccessNode('_')

    @staticmethod
    def expressions():
        return [
            nxutil.node_path_graph(RedundantArray._in_array,
                                   RedundantArray._out_array),
        ]

    @staticmethod
    def can_be_applied(graph, candidate, expr_index, sdfg, 
                       strict=False):
        in_array =  \ 
          graph.nodes()[candidate[RedundantArray._in_array]]
        out_array = \
          graph.nodes()[candidate[RedundantArray._out_array]]

        # Ensure out degree is one (only one target, out_array)
        if graph.out_degree(in_array) != 1:
            return False

        # Make sure that the candidate is a transient variable
        if not in_array.desc(sdfg).transient:
            return False

        # Make sure both arrays are using the same storage location
        if (in_array.desc(sdfg).storage !=  
              out_array.desc(sdfg).storage):
            return False

        # Find occurrences in this and other states
        occurrences = []
        for state in sdfg.nodes():
            occurrences.extend([
                n for n in state.nodes() 
                if isinstance(n, nodes.AccessNode)
                and n.desc(sdfg) == in_array.desc(sdfg)
            ])

        if len(occurrences) > 1:
            return False

        # If strict, only apply if arrays are of same shape 
        # (no need to modify memlet subset)
        if (strict and (len(in_array.desc(sdfg).shape) != 
                len(out_array.desc(sdfg).shape))
                or any(i != o for i, o in zip(
                    in_array.desc(sdfg).shape,
                    out_array.desc(sdfg).shape))):
            return False

        return True

    def apply(self, sdfg):
        graph = sdfg.nodes()[self.state_id]
        in_array = gnode(RedundantArray._in_array)
        out_array = gnode(RedundantArray._out_array)

        for e in graph.in_edges(in_array):
            # Modify all incoming edges to point to out_array
            path = graph.memlet_path(e)
            for pe in path:
                if pe.data.data == in_array.data:
                    pe.data.data = out_array.data

            # Redirect edge to out_array
            graph.remove_edge(e)
            graph.add_edge(e.src, e.src_conn, out_array, 
                           e.dst_conn, e.data)

        # Finally, remove in_array node
        graph.remove_node(in_array)

# Register transformation in global registry
pm.Transformation.register_pattern(RedundantArray)
\end{lstlisting}

\section{Graph Datasets}

Table \ref{tbl:graphtypes} lists the graph datasets used in the BFS case study of the paper.

\begin{table}[h]
	\caption{Graph Properties}\vspace{-1em}
	\label{tbl:graphtypes}
	\centering
	\scriptsize
	\begin{tabular}{|l|c|c|c|c|c|}
		\hline
		\bf    Name & \bf Nodes & \bf Edges & \bf Avg.   & \bf Max & \bf Size \\
		&           &           & \bf Degree & \bf Degree & \bf  (GB)\\\hline 
		\multicolumn{6}{|c|}{\bf Road Maps}\\\hline
		USA \cite{dimacs9}         & 24M & 58M & 2.41 & 9 & 0.62 \\\hline
		OSM-eur-k \cite{karlsruhe}   & 174M & 348M & 2.00 & 15 & 3.90 \\\hline
		\multicolumn{6}{|c|}{\bf Social Networks}\\\hline
		soc-LiveJournal1 \cite{soclj} & 5M & 69M & 14.23 & 20,293 & 0.56 \\\hline
		twitter \cite{cha2010twitter} & 51M  & 1,963M  & 38.37 & 779,958 & 16.00 \\\hline
		\multicolumn{6}{|c|}{\bf Synthetic Graphs}\\\hline
		kron21.sym \cite{dimacs10} & 2M & 182M  & 86.82 & 213,904 & 1.40 \\\hline
	\end{tabular}
\end{table}

\section{Indirect Memory Access}

\begin{figure}[h]
	\includegraphics[width=\linewidth,page=5]{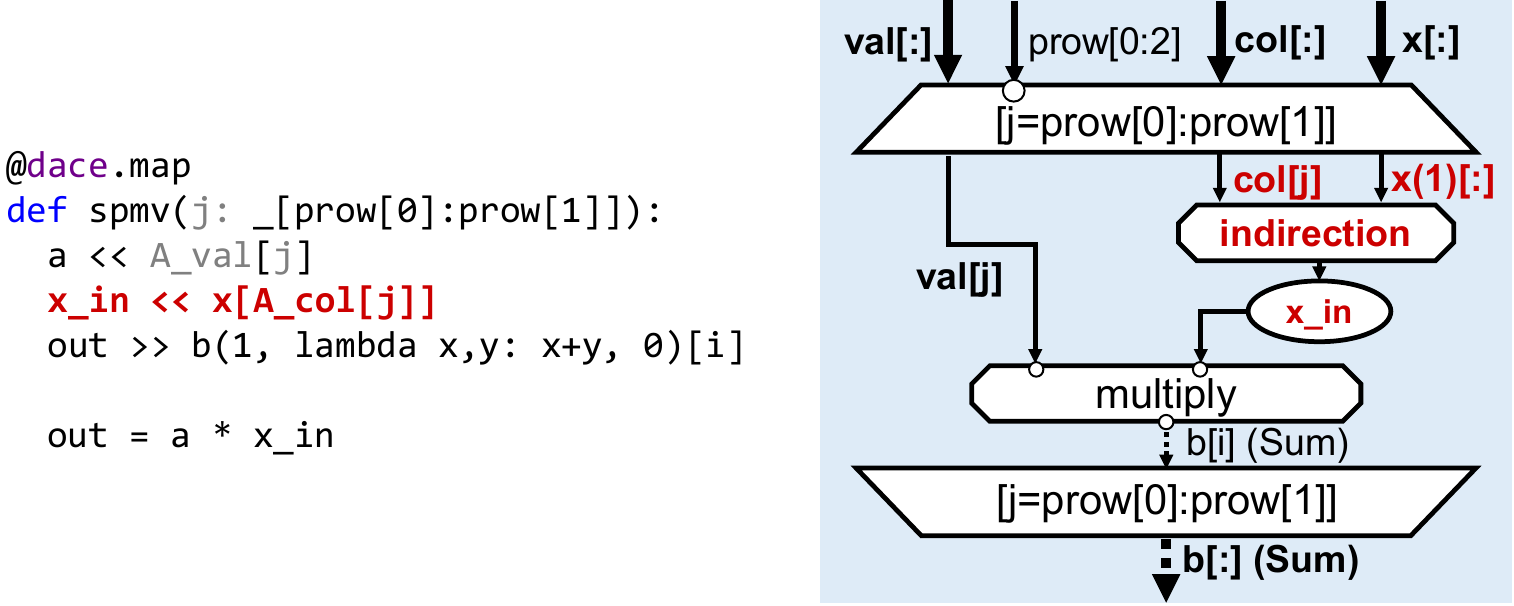}
	\vspace{-1em}
	\caption{Indirect Memory Access Dataflow}
	\label{fig:indirect2}
\end{figure}

Indirect memory access, i.e., \texttt{A[b[i]]}, is an important characteristic
of sparse data structure and pointer jumping algorithms.
Indirection cannot be directly represented by a single memlet. As shown in Fig.~\ref{fig:indirect2} (an excerpt of Sparse Matrix-Vector Multiplication), such statements are converted to a subgraph,
where the internal access is given exactly (\texttt{A\_col[j]}
in the figure), and the indirect memory is given by a memlet
with one access (\texttt{x(1)[:]}). The memory is then accessed dynamically
and copied using a tasklet.

\end{document}